\documentclass{article}

\usepackage{microtype}
\usepackage{graphicx}
\usepackage{subcaption}
\usepackage{booktabs} % for professional tables

\usepackage{hyperref}
\usepackage{caption}
\usepackage{multirow}
\usepackage{mathtools}
\usepackage{caption} 
\usepackage{cuted}
\usepackage{arydshln}
\usepackage{bm}
\usepackage[table]{xcolor}
\usepackage{colortbl}
\usepackage{tikz}
\usetikzlibrary{patterns}
\usepackage{pifont}
\definecolor{bestgreen}{RGB}{230,245,230}
\definecolor{secondblue}{RGB}{230,235,250}
\definecolor{lightblue}{RGB}{235, 245, 255}
\definecolor{lightgreen}{RGB}{230, 245, 230}
\newcommand{\cmark}{\ding{51}}
\newcommand{\xmark}{\ding{55}} % ✗

\newcommand{\best}[1]{\cellcolor{bestgreen}\(\bm{#1}\)}
\newcommand{\second}[1]{\cellcolor{secondblue}\underline{#1}}
\newcommand{\greentag}[1]{\colorbox{bestgreen}{\strut\textbf{#1}}}
\newcommand{\bluetag}[1]{\colorbox{secondblue}{\strut\underline{#1}}}
\usepackage[accepted]{icml2026}

\usepackage{amsmath}
\usepackage{amssymb}
\usepackage{amsthm}
\usepackage{adjustbox}
\usepackage{tabularx}
\usepackage{array}

\usepackage[capitalize,noabbrev]{cleveref}

\theoremstyle{plain}

\theoremstyle{definition}

\theoremstyle{remark}

\usepackage{url}
\usepackage{amsmath}
\usepackage{enumitem}
\usepackage{multirow}
\usepackage{float}
\usepackage{tcolorbox}
\tcbuselibrary{breakable}

\usepackage[textsize=tiny]{todonotes}

\icmltitlerunning{Training Diffusion Language Models for Black-Box Optimization}

\begin{document}

\twocolumn[
\icmltitle{Training Diffusion Language Models for Black-Box Optimization}

% It is OKAY to include author information, even for blind
% submissions: the style file will automatically remove it for you
% unless you've provided the [accepted] option to the icml2025
% package.

% List of affiliations: The first argument should be a (short)
% identifier you will use later to specify author affiliations
% Academic affiliations should list Department, University, City, Region, Country
% Industry affiliations should list Company, City, Region, Country

% You can specify symbols, otherwise they are numbered in order.
% Ideally, you should not use this facility. Affiliations will be numbered
% in order of appearance and this is the preferred way.
\icmlsetsymbol{equal}{*}
\icmlsetsymbol{lead}{\textdagger}
\icmlsetsymbol{indep}{\ddag}

\begin{icmlauthorlist}
\icmlauthor{Zipeng Sun}{equal,mcgill,mila}
% \icmlauthor{Can (Sam) Chen}{equal,lead,mila,amazon} 
\icmlauthor{Can (Sam) Chen}{equal,lead,indep,mila,amazon} 
\icmlauthor{Ye Yuan}{mcgill,mila}
\icmlauthor{Haolun Wu}{mcgill,mila}
\icmlauthor{Jiayao Gu}{mcgill,mila}
\icmlauthor{Chris Pal}{mila,poly,cifar}
\icmlauthor{Xue Liu}{mcgill,mila,mbzuai}
\end{icmlauthorlist}

\icmlaffiliation{mcgill}{McGill University}
\icmlaffiliation{mila}{MILA - Quebec AI Institute}
\icmlaffiliation{poly}{Polytechnique Montreal}
\icmlaffiliation{cifar}{Canada CIFAR AI Chair}
\icmlaffiliation{mbzuai}{Mohamed bin Zayed University of Artificial Intelligence}
\icmlaffiliation{amazon}{Amazon AGI}

\icmlcorrespondingauthor{Zipeng Sun}{zpointsun@gmail.com}
\icmlcorrespondingauthor{Can (Sam) Chen}{chencan421@gmail.com}

% You may provide any keywords that you
% find helpful for describing your paper; these are used to populate
% the "keywords" metadata in the PDF but will not be shown in the document
% \icmlkeywords{Machine Learning, ICML}

\vskip 0.3in
]

% this must go after the closing bracket ] following \twocolumn[ ...

% This command actually creates the footnote in the first column
% listing the affiliations and the copyright notice.
% The command takes one argument, which is text to display at the start of the footnote.
% The \icmlEqualContribution command is standard text for equal contribution.
% Remove it (just {}) if you do not need this facility.

%\printAffiliationsAndNotice{}  % leave blank if no need to mention equal contribution
\printAffiliationsAndNotice{\icmlEqualContribution\ \textdagger Project lead. \ddag Work done independent of the author's position at Amazon AGI.}
% \printAffiliationsAndNotice{\icmlEqualContribution} 
% otherwise use the standard text.

\begin{abstract}

We study offline black-box optimization (BBO), aiming to discover improved designs from an offline dataset of designs and labels, a problem common in robotics and DNA with limited labeled samples.
While recent work applies autoregressive LLMs to BBO by formatting tasks as natural-language prompts, their left-to-right design generation struggles to capture the strong bidirectional dependencies inherent in design problems. 
To address this, we propose adapting diffusion LLMs to offline BBO to leverage their bidirectional modeling capabilities. 
However, a domain gap exists between the natural text pre-training of diffusion LLMs and the heterogeneous signals in BBO (prompts, designs, and labels). 
To bridge this gap, we construct a unified prompt–-response corpus and introduce delimiter tokens to explicitly mark field boundaries for \textit{domain adaptation}.
We further propose a two-stage \textit{post-training} framework to align the diffusion LLM generation with high-label designs. 
The first stage performs supervised fine-tuning on the unified dataset via masked-response prediction, and the second stage adopts reinforcement learning with rewards defined by label improvements.
Our method achieves state-of-the-art results on Design-Bench under small-data settings with highly efficient training, requiring only $1.5$ H100 GPU hours for discrete tasks. Code for our work is available 
% \href{https://anonymous.4open.science/r/Anonymous-dllm4bbo-D78A/README.md}{here}.
\href{https://github.com/zpointS/DiBO}{here}.

% \ZP{TODO: Update public repo.}

\end{abstract}

% {
% \setlength{\textfloatsep}{1pt}
\begin{figure*}[t]
  % \captionsetup{skip=15pt}
  % \vspace{-4mm}
  %   \setlength{\abovecaptionskip}{2pt}
  % \setlength{\belowcaptionskip}{0pt}
  \centering
  \includegraphics[width=\linewidth]{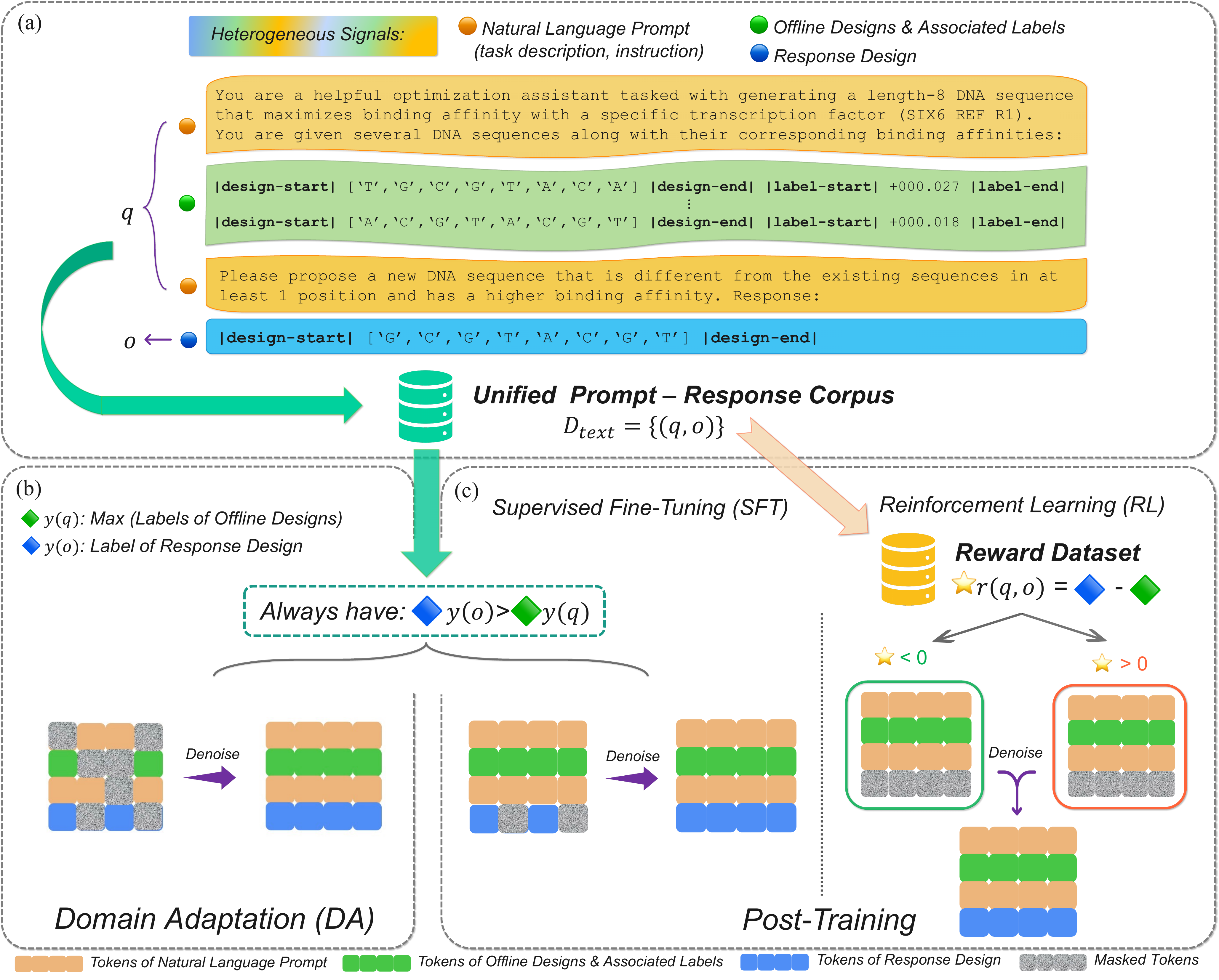}
    \caption{
    \textbf{Overview of the DiBO framework.} 
    (a) Unified Prompt–Response Corpus: Heterogeneous BBO signals (natural-language prompts, offline designs and their associated labels) are unified using explicit delimiter tokens.
    (b) Domain Adaptation (DA): The diffusion LLM is domain-adapted via joint masked-token prediction over prompts and responses.
    (c) Post-Training: 
    The model is further aligned with high-label designs via supervised fine-tuning (SFT) and reinforcement learning (RL) based on label improvements.
    }
  \label{fig:main_fig}
  \vspace{-4mm}
\end{figure*}
% }
\vspace{-5mm}
% \vspace{-2mm}
% \vskip -20pt

\section{Introduction}
Across diverse scientific disciplines, including robotic design, DNA synthesis, and materials discovery, researchers aim to create novel designs that maximize performance metrics~\citep{trabucco2022design}. 
Nevertheless, obtaining these property measurements often requires direct experimentation that is both labor-costly and time-intensive~\citep{hamidieh2018data,angermueller2019model,barrera2016survey,sample2019human}. 
This constraint essentially precludes the use of standard iterative online optimization. 
As an alternative, the community turns to offline black-box optimization (BBO)~\citep{kim2026offline}, which operates by utilizing an existing, static dataset of design-property pairs to propose superior candidates. 
The central difficulty in this paradigm is that real-world applications frequently suffer from the lack of labeled data points.

Traditional approaches often rely on training task-specific surrogate models or generative models, but they suffer from epistemic uncertainty arising from the aforementioned limited data coverage~\citep{kim2026offline}.
More recent work reformulates task descriptions and offline datasets as natural-language prompts, enabling LLMs to directly generate candidate designs~\citep{yang2023large, zhang2023using, liu2024large, nie2024importance, velivckovic2024amplifying, novikov2025alphaevolve}.
However, most of these methods employ autoregressive (AR) LLMs~\citep{brown2020language}, which are inherently unidirectional. 
Many design problems (e.g., DNA sequence generation) exhibit bidirectional dependencies, where each unit can be influenced by both its prefix and suffix, making AR models insufficient to fully capture these interactions during left-to-right generation.
% However, most of these methods employ autoregressive (AR) LLMs~\citep{brown2020language}, which are inherently unidirectional.
% Many design problems—such as DNA sequence generation—exhibit bidirectional dependencies, where each unit can be influenced by both its prefix and suffix.
% This unidirectional factorization forces AR models to commit to early decisions without access to future context, leading to suboptimal modeling of structural dependencies and limiting their effectiveness in complex design spaces.

In response, we turn to diffusion LLMs for BBO, leveraging their inherent bidirectional modeling capabilities.
A diffusion LLM is trained by progressively masking tokens and learning to reconstruct them, and at inference time, it reverses this process by iteratively denoising a fully masked sequence~\citep{nie2025large, gong2025diffucoder, arriola2025block}.
This training paradigm naturally enables diffusion LLMs to capture bidirectional dependencies.

% \ZP{Compare with ICLR submission (other works using diffusion) in this paragraph? Not sure how to write this paragraph}
% Our work differs from prior diffusion-based optimization methods that focus on modeling optimization trajectories
% or rely on explicit surrogate objectives.
% %
% Instead, we study how to directly align a pre-trained diffusion LLM with offline BBO through training-time adaptation,
% using only a static dataset of design--label pairs.
% %
% Compared to recent LLM-based optimizers that emphasize test-time search or inference-time scaling,
% our approach focuses on representation alignment and learning to generate improved designs
% within a single forward denoising step.

As illustrated in Figure~\ref{fig:main_fig}(a), the BBO setting naturally involves multiple heterogeneous signals: 
natural-language prompts (task descriptions and instructions) together with offline designs and their associated labels.
This introduces a domain gap, as the diffusion LLM is pretrained solely on natural-language text.
To bridge this gap, we construct a unified prompt–response corpus and extend the tokenizer 
with delimiter tokens—(1)\texttt{|design-start|}/\texttt{|design-end|} to enclose designs and 
(2)\texttt{|label-start|}/\texttt{|label-end|} to enclose labels, as shown in the middle part of Figure~\ref{fig:main_fig} (a).
These delimiters explicitly mark the semantic roles of text, design, and label within a unified input sequence.
We then perform \textit{domain adaptation} (Figure~\ref{fig:main_fig}(b)), 
optimizing the diffusion LLM to jointly predict masked tokens in both the prompt and the response.
% \ZP{Should we introduce the kernel distance based neighbor selection here?}

We further propose a two-stage \textit{post-training} framework illustrated in Figure~\ref{fig:main_fig}(c), 
to align the diffusion LLM’s response generation with high-label designs. 
In the first stage, we perform supervised fine-tuning~(SFT) on the unified dataset 
via masked-response prediction, which instills an inductive bias towards high-label regions of the design space.
In the second stage, we construct a reinforcement learning (RL) dataset 
where rewards are defined as label improvements from prompt to response,
and employ an efficient one-step log-probability approximation to compute the RL loss.
This stage further integrates fine-grained reward feedback into the diffusion LLM.

% \vspace{2mm}
% \ZP{Remember to erase the vspace after main figure.}
We summarize our main contributions as follows:
% \begin{itemize}[leftmargin=*]
\begin{itemize}[leftmargin=*, itemsep=0.5em, topsep=0em, parsep=0em, partopsep=0em]
    \item We propose adapting diffusion LLMs to offline BBO to exploit their bidirectional modeling, and construct a unified prompt--response corpus with semantic delimiters to facilitate \textit{domain adaptation}.

    \item We further propose a two-stage \textit{post-training} framework using SFT and RL to effectively align diffusion generation with high-label designs.

    \item We demonstrate state-of-the-art performance on Design-Bench in small-data settings.
\end{itemize}
% \begin{itemize}[leftmargin=*]
% We propose adapting diffusion LLMs to offline BBO to exploit their bidirectional modeling, and construct a unified prompt--response corpus with semantic delimiters to facilitate \textit{domain adaptation}.

% We further propose a two-stage \textit{post-training} framework using SFT and RL to effectively align diffusion generation with high-label designs.

% We demonstrate state-of-the-art performance on Design-Bench in small-data settings.
% \end{itemize}

\section{Preliminaries and Related Work}
% \begin{figure*}[ht]
%   \centering
%   \includegraphics[width=\linewidth]{figures/DiBO.pdf}
%   \caption{Method overview of DiBO.}
%   \label{fig:main_fig}
% \end{figure*}

% \ZP{Please help me check if the statements / formulas are correct in this section.}

\subsection{Offline Black-Box Optimization}

Offline Black-Box Optimization (BBO) aims to identify an optimal design $\boldsymbol{x}^{*} \in \mathcal{X} \subseteq \mathbb{R}^{D}$ that maximizes an unknown objective function $f: \mathcal{X} \rightarrow \mathbb{R}$ associated with the design $\boldsymbol{x}$:
\begin{equation}
\label{eq:bbo-objective}
\boldsymbol{x}^{*} = \arg\max_{\boldsymbol{x} \in \mathcal{X}} f(\boldsymbol{x}).
\end{equation}
A practical example involves identifying a DNA sequence to maximize its binding affinity with a specific protein. 
In many practical scenarios, direct interaction with the objective function is prohibitively expensive or time-consuming. Consequently, we operate under the assumption that we only have access to an offline dataset of previously labeled designs $\mathcal{D} = \{(\boldsymbol{x}_i, y_i)\}_{i=1}^{N}$, where $y_i = f(\boldsymbol{x}_i)$ \citep{trabucco2022design, kim2026offline}. 
Our work specifically focuses on small-data settings where the number of labeled samples $N$ is limited (e.g., $N=500$), reflecting the scarcity of labeled data in real-world applications.

A common baseline is to fit a surrogate model — such as a Deep Neural Network or a Gaussian Process — to approximate $f(\cdot)$ in a supervised manner and then leverage it to guide the design optimization. However, this approach often suffers from the {out-of-distribution} issue, where the optimizer explores regions of the design space that the surrogate model cannot accurately predict. 
In this paper, we propose to explore diffusion LLMs due to their robust generalization and bidirectional modeling capabilities.

\subsection{Diffusion Large Language Models}

While autoregressive (AR) models such as GPT \citep{brown2020language} generate text in a left-to-right way, diffusion LLMs \citep{nie2025large, gong2025diffucoder} provide a non-causal framework by modeling the data distribution through iterative refinement. This approach offers the distinct advantage of global, bidirectional modeling. Such capabilities are essential in design domains where functional dependencies are often non-sequential; for example, a specific unit within a DNA sequence is constrained not only by its prefix but also by its suffix.

Diffusion LLMs characterize the distribution $p_{\boldsymbol{\theta}}(\boldsymbol{x}_0)$ via two symmetric processes: forward corruption and reverse reconstruction. Given an initial sequence $\boldsymbol{x}_0$, the forward process introduces noise by stochastically replacing tokens with a \texttt{[MASK]} symbol according to a timestep $t \sim \mathcal{U}[0,1]$. This yields a partially corrupted sequence $\boldsymbol{x}_t$, which becomes fully masked as $t \to 1$. 

The model optimizes a mask predictor $p_{\boldsymbol{\theta}}(\cdot \mid \boldsymbol{x}_t)$ to recover the original tokens from their corrupted states. The training objective is defined as:
\begin{equation}
% \small
% \footnotesize
\label{eq:dllm-loss}
\mathcal{L}_{\text{dLLM}} = - \mathbb{E} \left[ 
    \frac{1}{t} \sum_{i=1}^{L} 
    \mathbf{1}[x^{i}_{t} = \texttt{[M]}] 
    \log p_{\boldsymbol{\theta}}(x^{i}_{0} \mid \boldsymbol{x}_t) 
\right],
\end{equation}

% 
% where $t$ is sampled uniformly from the interval $[0, 1]$, \texttt{[M]} is the \texttt{[MASK]} token, $L$ is the sequence length, and $\boldsymbol{\theta}$ denotes the parameters of the diffusion LLM. Once trained, the model generates sequences by starting from total occlusion and iteratively denoising the sequence over a predefined schedule.
where $t$ is sampled uniformly from the interval $[0, 1]$, \texttt{[M]} denotes the \texttt{[MASK]} token used for masking, $L$ is the sequence length, and $\boldsymbol{\theta}$ represents the parameters of the diffusion LLM. 
% The training objective focuses exclusively on masked positions, enabling the model to learn conditional token distributions given arbitrary partial observations.
Once trained, the model generates sequences by starting from total occlusion and iteratively denoising the sequence over a predefined schedule, progressively refining the sequence until convergence.

\subsection{LLMs for Black-Box Optimization}

The expressive capacity of LLMs has recently been leveraged to advance Black-Box Optimization (BBO)~\citep{song2024position}. Current research generally bifurcates into two paradigms: (1) \textbf{Predictors}, which treat LLMs as surrogate models fine-tuned to estimate the objective value $y$ for a given design $\boldsymbol{x}$~\citep{raffel2020exploring, nguyen2024predicting, tan2025towards}; and (2) \textbf{Generators}, which utilize the generative priors of LLMs to directly sample candidate designs $\boldsymbol{x}$ via task-specific prompting \citep{zhang2023using, liu2024large, velivckovic2024amplifying, novikov2025alphaevolve}.

Within the generative paradigm, autoregressive (AR) LLMs are typically favored, using task descriptions and collected designs as context for design sampling.
For instance, by providing a prompt such as \textit{``Design a DNA sequence with high binding affinity''} alongside offline data, the LLM can be prompted to generate optimized sequences as a response.

% \begin{figure}[t]
% \vspace{-0.6em}
% \centering
% \begingroup
% \setlength{\fboxsep}{2.5pt}   % tighten padding
% \setlength{\fboxrule}{0.4pt}
% \scriptsize                    % smaller font
% \fbox{%
% \begin{minipage}{0.95\linewidth}
% \ttfamily
% % \textbf{Prompt:}\par
% Design a DNA sequence with high binding affinity.\par
% \textbf{Offline data:}\par
%  [\textquotesingle C\textquotesingle,\textquotesingle C\textquotesingle,\textquotesingle A\textquotesingle,\textquotesingle C\textquotesingle,\textquotesingle A\textquotesingle,\textquotesingle C\textquotesingle,\textquotesingle G\textquotesingle,\textquotesingle T\textquotesingle];
%  0.018 
%  [\textquotesingle T\textquotesingle,\textquotesingle C\textquotesingle,\textquotesingle G\textquotesingle,\textquotesingle C\textquotesingle,\textquotesingle T\textquotesingle,\textquotesingle A\textquotesingle,\textquotesingle G\textquotesingle,\textquotesingle G\textquotesingle]
% \;
% 0.027\par
% \textbf{Instruction:} Propose a new high-affinity design.\par
% \textbf{Response:}\par
%  [\textquotesingle A\textquotesingle,\textquotesingle C\textquotesingle,\textquotesingle G\textquotesingle,\textquotesingle A\textquotesingle,\textquotesingle G\textquotesingle,\textquotesingle A\textquotesingle,\textquotesingle G\textquotesingle,\textquotesingle T\textquotesingle]

% \end{minipage}%
% }
% \endgroup
% \vspace{-0.8em}
% % \caption{Example.}
% \label{fig:prompt_example}
% \vspace{-0.9em}
% \end{figure}

While our work follows this generative lineage, we diverge in two fundamental aspects. First, whereas prior efforts rely on AR models constrained by unidirectional causal masking~\citep{zhang2023using, liu2024large}, we employ {diffusion LLMs} to exploit their inherent bidirectional modeling. This global context is better suited for capturing the intricate structural dependencies common in complex design spaces. Second, we transition from pure prompting to {model adaptation}. While extant diffusion LLM research~\citep{yuan2026diffusion} is often restricted to frozen-model prompting in few-shot settings, we propose \textit{domain adaptation} and \textit{post-training} on these models. This allows the diffusion LLMs 
% \Ye{this abbreviation is used before defining it?} 
to internalize domain-specific constraints and navigate the optimization landscape effectively, even in data-sparse regimes.

\section{Methodology}

\subsection{Domain Adaptation}

\paragraph{Heterogeneous Signals}
The BBO setting inherently involves heterogeneous inputs: natural-language prompts (task descriptions and instructions) alongside offline designs and their corresponding labels.
Since diffusion LLMs are pretrained on natural text, they would otherwise treat designs and labels as ordinary text, leading to ineffective representations and a domain gap.
To address this problem, we extend the tokenizer with delimiter tokens: (1) \texttt{|design-start|}/\texttt{|design-end|} to mark designs, and (2) \texttt{|label-start|}/\texttt{|label-end|} to enclose label values. These delimiters explicitly define the semantic roles of text, design, and label components within a unified input sequence.

% \ZP{This "Unified" corpus is only "directly" used in da  and sft. In rl we used "reward dataset". I'm afraid this "unfied" is somehow misleading. I'm not sure if this is a problem.}

\paragraph{Unified Corpus}
We construct a unified prompt--response corpus that serves as the foundation for both domain adaptation and the subsequent supervised fine-tuning stage.
Each input prompt $q$ is formed by concatenating: (1) a task description specifying the semantics, format, and optimization objective of the design and its associated label, and (2) an offline dataset consisting of design–label pairs, followed by an instruction to generate improved designs. The corresponding response $o$ is a design whose label value exceeds all those in the prompt. 
An example of the final prompt--response format can be found in Figure~\ref{fig:main_fig}(a).
Detailed construction of the dataset $\mathcal{D}_{text}=\{(q, o)\}$ is provided in Appendix~\ref{app: offline-data-construction} and ~\ref{app: prompt-response-construction}. Notably, to facilitate effective reasoning by the diffusion LLM, the target response should remain within a reasonable distribution of the prompt examples. 
% We define the neighbor of a given response design $o$ as
Consequently, we select prompt designs based on pairwise kernel similarity to the response,
and we elaborate the details in Appendix~\ref{app: design-similarity}.

\paragraph{Joint Prompt--Response Loss}
We adopt a joint loss that reconstructs masked tokens in both the prompt $q$ and the response $o$ for domain adaptation:

\vspace{-4mm}

{ 
\small
% \scriptsize
\begin{equation}
\label{eq:jointloss}
\mathcal{L}_{\text{DA}}
= - \mathbb{E}\!\left[
  \frac{1}{t} \sum_{i=1}^{L}
  \mathbf{1}[q^{i}_{t} = \texttt{[M]}, o^{i}_{t} = \texttt{[M]}]
  \log p_{\boldsymbol{\theta}}(q^{i}_{0}, o^{i}_{0} \mid q_{t}, o_{t})
\right],
\end{equation}
}
% \begin{equation}
% \label{eq:jointloss}
% {
% % \small
% \begin{aligned}
% \mathcal{L}_{\text{DA}}
% &= - \mathbb{E}\!\Bigg[
%   \frac{1}{t} \sum_{i=1}^{L}
%   \mathbf{1}\!\left[o^{i}_{t} = \texttt{[MASK]}\right]
% \\
% &\qquad\qquad\cdot
%   \log p_{\boldsymbol{\theta}}\!\left(q^{i}_{0}, o^{i}_{0} \mid q_{t}, o_{t}\right)
% \Bigg].
% \end{aligned}
% }
% \end{equation}
%
where $t$ is sampled uniformly from the interval $[0, 1]$, $L$ is the sequence length, 
$q_{0}, o_{0}$ are clean samples from $\mathcal{D}_{\text{text}}$, 
$q_{t}, o_{t}$ are their masked counterparts, and $\boldsymbol{\theta}$ denotes the parameters of the diffusion LLM.
This joint objective enables the model to learn the semantic roles of the delimiter tokens and adapt effectively to the BBO domain context.

\subsection{Post-Training}

We introduce a two-stage \textit{post-training} framework to align diffusion LLMs with high-label design generation. 

\paragraph{Supervised Fine-Tuning}
We reuse the unified prompt--response corpus but optimize only over the response sequences, using the following objective:
\begin{equation}
\mathcal{L}_{\text{SFT}}
= - \mathbb{E}\!\left[
  \frac{1}{t}
  \sum_{i=1}^{L}
  \mathbf{1}[o^{i}_{t} = \texttt{[M]}]
  \log p_{\boldsymbol{\theta}}(o^{i}_{0} \mid q_{0}, o_{t})
\right].
\end{equation}
%
% where $t$ is sampled uniformly from $[0, 1]$, $L$ is the sequence length, 
% $q_{0}, o_{0}$ are clean samples from $\mathcal{D}_{\text{text}}$, 
% $o_{t}$ is the masked version of $o_{0}$, and $\boldsymbol{\theta}$ denotes the parameters of the diffusion LLM.
%
This supervised fine-tuning stage provides a simple and stable alignment signal that encourages the diffusion LLM to generate improved designs conditioned on the prompt.
By reconstructing only masked response tokens, the model learns to map the few-shot prompt context to high-label design outputs.
This serves as an effective initialization for the subsequent reinforcement learning stage, which further incorporates fine-grained reward feedback.

\paragraph{Reward Dataset}
We then construct a reinforcement learning dataset $\mathcal{D}_{{rl}}$, where rewards are defined as label improvements from the prompt to the response:
\begin{equation}
    r(q, o) = y(o) - y(q),
\end{equation}
where $y(q)$ denotes the highest label among designs appearing in the few-shot prompt.
Unlike in SFT, where the response is always constructed to outperform the prompt and thus yields non-negative rewards, the RL phase imposes no such constraint.
%
% This construction introduces both positive and negative reinforcement signals while preserving local semantic similarity between prompt and response designs, as detailed in Appendix~\ref{ref:app_prompt_response_construction}.
This construction introduces both positive and negative rewards while preserving local semantic similarity between prompt and response designs, as detailed in Appendix~\ref{app: prompt-response-construction} and \ref{app: prompt-templates}.

\paragraph{Reinforcement Learning}
For each pair $(q, o)$, the log probability required by the RL objective is approximated using a one-step unmasking strategy~\citep{zhao2025d1} instead of iterative denoising for efficiency and stability:
\begin{equation}
    \log p_{\boldsymbol{\theta}}(o \mid q)
    \approx \sum_{k=1}^{|o|}
    \log p_{\boldsymbol{\theta}}(o_k \mid q, o_{\text{fullmask}}),
\end{equation}
where $o_k$ is the $k$-th token of the response and $o_{\text{fullmask}}$ denotes a fully masked response sequence.

\textcolor{black}{
Since the designs in offline BBO are collected from external real-world processes rather than generated by a known language model, the behavior policy (i.e., the old policy) that produced those designs is inaccessible.
Following~\citet{yan2025learningreasonoffpolicyguidance, fu2025srft}, we therefore assume this old policy to be uniform.
Under this assumption, the denominator of the importance-sampling ratio becomes a constant, and ratio clipping and KL regularization are not applied.
}
The resulting RL objective is:
\begin{equation}
    \mathcal{L}_{\text{RL}} 
    = - \mathbb{E}_{q, o}  \left[ \frac{1}{|o|}
      \sum_{k=1}^{|o|} p_{\boldsymbol{\theta}}(o_k \mid q, o_{\text{fullmask}} ) \frac{r(q, o)}{\sigma} 
    \right],
\end{equation}
where $(q, o)$ are sampled from $\mathcal{D}_{{rl}}$, and $\sigma$ denotes the standard deviation of rewards. 
In this formulation, the term $\frac{r(q, o)}{\sigma}$ serves as the advantage signal. We intentionally omit reward centering (i.e., subtracting the mean) to preserve the inherent prompt-specific information. 
The RL optimization integrates fine-grained reward feedback into the diffusion LLM, in contrast to the coarse binary supervision in SFT.

% TODO: compare the uniform policy with the SFT policy assumption. In that case, we need add a clip ratio.

\section{Experiments}
We conduct comprehensive experiments to evaluate the effectiveness of our method in offline black-box optimization,
with benchmarks detailed in Section~\ref{subsec: benchmark},
baselines in Section~\ref{subsec: baseline},
implementation details in Section~\ref{subsec: implementation_detail},
results in Section~\ref{subsec: result_and_analysis},
ablations in Section~\ref{subsec: ablation},
and hyperparameter settings described in Section~\ref{subsec: hyper}.

\subsection{Benchmarks}
\label{subsec: benchmark}

\paragraph{Datasets}
%\ZP{I suppose we should change the TF-10 description according to how ye changed its data.}
We consider two discrete sequence design problems and two continuous-parameter tasks drawn from the Design-Bench suite~\citep{trabucco2022design}.
The \emph{discrete} tasks are:
\textbf{(1) TF Bind 8 (TF8)}~\citep{barrera2016survey}, which requires designing an $8$-length DNA sequence to maximize binding activity with the \texttt{SIX6\_REF\_R1} transcription factor; and
% \ZP{Ye please help check the description and the cite of TF-10.}
% \Ye{please cite this paper for TF10: https://www.pnas.org/doi/10.1073/pnas.1715888115}
% \ZP{Check tf10 here.}
% \textbf{(2) TF Bind 10 (TF10)}~\cite{doi:10.1073/pnas.1715888115}, an extension to a $10$-length sequence with the same objective.
% \Ye{We may not say this is an extension, because they are from different authors.}
% %
% Note: Upon closer inspection, we found the original TF Bind 10 dataset was formulated to minimize DNA-protein binding affinity. To align this with a standard maximization objective, we negated the values and reorganized the data accordingly.
% \Ye{I would say design bench mistakenly use ddG as the labels. We correct them to binding affinity from the original paper to align with the task description in design-bench.}
\textbf{(2) TF Bind 10 (TF10)}~\citep{doi:10.1073/pnas.1715888115}, a $10$-length DNA sequence design task that optimizes transcription factor binding affinity.
%
% Note: Upon closer inspection, we found the original TF Bind 10 labels corresponded to binding free energy difference (ddG), which was formulated to minimize the binding affinity; we negate the ddG values from the original paper \cite{doi:10.1073/pnas.1715888115} to align with the task description in Design-Bench.
Note: The labels of TF Bind $10$ in Design-Bench correspond to binding free energy difference (ddG), where lower values indicate stronger binding; we therefore negate the ddG values from the original paper~\citep{doi:10.1073/pnas.1715888115} so that higher scores correspond to stronger binding, consistent with the Design-Bench task description.
% Note: The original TF Bind 10 labels correspond to binding free energy (ddG) and are formulated as a minimization objective; we negate the ddG values so that higher scores indicate stronger binding, consistent with the Design-Bench task description.
%
The \emph{continuous} tasks are:
\textbf{(3) Ant Morphology (Ant)}~\citep{brockman2016openai}, which optimizes a $60$-dimensional ant body morphology for fast crawling;
and \textbf{(4) D’Kitty Morphology (D’Kitty)}~\citep{ahn2020robel}, 
which optimizes a $56$-dimensional D’Kitty robot morphology to navigate toward a fixed target location.
% which optimizes a $56$-D D’Kitty robot morphology for the same objective; 
% \Ye{D'Kitty is to navigate the robot to a fixed location.}
% and \textbf{(5) Superconductor}~\cite{hamidieh2018data,trabucco2022design}, which aims to design a superconducting material with a high critical temperature, represented by an $86$-dimensional continuous vector encoding its chemical composition.
% Perhaps include a data-mixture optimization task here, which is a mixed discrete and continuous task.
% For all tasks, we construct few-shot settings by uniformly sampling $\mathbf{9}$ labeled examples from the offline dataset to form in-context prompts and training pairs (10 samples may induce out-of-memory error on a single H100 80G GPU).

\vspace{-3mm}

\paragraph{Evaluation}
For all methods, we use the task-specific oracle from the \textit{Design-Bench Benchmark Tasks}~\citep{trabucco2022design} to score generated designs.
Following standard practice~\citep{trabucco2021conservative}, each method is allowed to propose $128$ candidates per task, and we use the best-achieved value (i.e., the $100^{\text{th}}$ percentile) of the normalized ground-truth score as the main metric.
The normalized score $y_n$ is defined as $y_{n} = \frac{y - y_{\min}}{y_{\max} - y_{\min}}$,
where $y$ denotes the raw oracle score of a design, and $y_{\min}$, $y_{\max}$ are the minimum and maximum scores over the full (unobserved) dataset.
% For evaluation, we choose random examples in prompt context to boost exploitation, otherwise they seen combination already seen and lack of exploits.
% To make comparisons more robust, we additionally report the mean and median ranks of all methods across tasks, together with the median normalized scores.

\subsection{Baselines}
\label{subsec: baseline}

We systematically benchmark our approach against a wide range of prior methods.

\vspace{-2mm}

\paragraph{Forward Methods} We consider methods that utilize a learned surrogate to guide design optimization.
\textit{(1) Grad (mean)} models the black-box function with a Gaussian Process (GP) and performs gradient ascent on the posterior mean to improve existing designs.
\textit{(2) Grad (EI)} optimizes the Expected Improvement (EI) acquisition function instead of the mean. 
% \textcolor{red}{For GP baselines, \textit{Grad (mean)} and \textit{Grad (EI)}, we use the dimension-scaled GP prior configuration of \citet{hvarfner2024vanilla}, which is better suited to high-dimensional Bayesian optimization as discussed in Appendix~\ref{app: high-dim}.}
\textcolor{black}{For GP baselines, we use the dimension-scaled GP prior configuration of \citet{hvarfner2024vanilla}, which is better suited to high-dimensional Bayesian optimization as discussed in Appendix~\ref{app: high-dim}.}
% \ZP{TODO: Update main table results.}
\textit{(3) COMs}~\citep{trabucco2021conservative} lower-bounds a neural surrogate’s predictions on out-of-distribution inputs and applies gradient ascent on this conservative surrogate.
\textit{(4) ICT}~\citep{yuan2023importance} maintains three surrogate networks with a rotating pseudo-labeling and co-teaching scheme to mitigate overfitting.
\textit{(5) MATCH-OPT}~\citep{hoang2025learningsurrogatesofflineblackbox} explicitly bounds the mismatch between surrogate gradients and true gradients to improve the quality of surrogate-guided optimization.
% \textit{(6) UniSO-T}~\citep{tan2025towards} models the black-box function as an autoregressive sequence-to-sequence model and uses this learned proxy to steer design generation. \Ye{no single word}
\textit{(6) UniSO-T}~\citep{tan2025towards} models the black-box function as an autoregressive sequence-to-sequence model and uses this learned surrogate to steer design generation.

% \textcolor{red}{
%     For GP-based baselines (Grad-mean and Grad-EI), we follow recent recommendations for high-dimensional Bayesian optimization and instantiate the GP surrogate with the dimension-scaled prior configuration of \citet{hvarfner2024vanilla}. 
%     This replaces the standard RBF GP used in our initial experiments while keeping the rest of the baseline pipeline unchanged. 
%     We report the updated results in the main tables and provide an RBF-vs-high-dimensional comparison in Appendix~\ref{app:hd_gp}.
% }

\vspace{-2mm}

\paragraph{Inverse Methods} These methods learn a conditional distribution over high-quality designs using generative models.
\textbf{VAE:}
\textit{(1) CbAS}~\citep{brookes2019conditioning} fits a VAE to the offline data with high-scoring samples emphasized.
\textit{(2) ExPT}~\citep{nguyen2023expt} fits a Transformer-based VAE on offline data, followed by in-context optimization to sample improved designs.
\textbf{GAN:}
\textit{(3) MIN}~\citep{kumar2020model} employs a GAN to model the inverse mapping from scores to designs, conditioning on high scores for direct sampling.
\textbf{Autoregressive:}
\textit{(4) BONET}~\citep{krishnamoorthy2022generative} models the trajectory from low- to high-scoring samples with an autoregressive model and unrolls these trajectories at test time to produce new candidates.
\textit{(5) OPRO}~\citep{yang2023large} prompts an autoregressive LLM with sequences of past design-label pairs to directly generate new designs. We instantiate OPRO with LLaMA3-8B-Instruct~\citep{grattafiori2024llama}, whose scale is comparable to our diffusion LLM.
\textbf{Diffusion:}
\textit{(6) GTG}~\citep{yun2024guided} trains a conditional diffusion model on synthetic trajectories derived from offline data to guide designs toward high-score regions.
\textit{(7) DDOM}~\citep{krishnamoorthy2023diffusion} learns a conditional diffusion model on the offline dataset and employs classifier-free guidance during sampling.
% \textit{(8) dLLM}~\citep{krishnamoorthy2023diffusion} directly queries the dLLM with prompts and design-label pairs to generate high-scoring designs via MCTS without fine-tuning the model.
\textit{(8) dLLM}~\citep{yuan2026diffusion} directly queries a frozen diffusion LLM with prompts and design-label pairs to generate high-scoring designs via MCTS without fine-tuning the model.
%
% Finally, we include \textit{CMA-ES}~\citep{hansen2006cma}, an evolutionary baseline that models the design via a covariance matrix over the design space and samples candidates from there.
Finally, we include \textit{(9) MCTS-transfer}~\citep{wang2024monte}, which employs Monte Carlo Tree Search (MCTS) to adaptively explore subspaces for improved design generation,
and \textit{(10) CMA-ES}~\citep{hansen2006cma}, an evolutionary baseline that models the design via a covariance matrix over the design space and samples candidates from there.

\subsection{Implementation Details}
\label{subsec: implementation_detail}

\begin{table*}[ht]
	\centering
	\caption{
    % Experimental results in $100$-th percentile normalized scores on four tasks for comparison.
    \textbf{Main results on Design-Bench benchmarks.}
    We report the $100$-th percentile normalized oracle score over $128$ generated candidates on four tasks: Ant Morphology, D’Kitty Morphology, TF Bind $8$, and TF Bind $10$.
    Results are averaged over $8$ random seeds (mean ± std).
    % Mean Score is the average score per method across 4 tasks, and 
    Rank Mean and Rank Median denote the average and median rank across all tasks.
    The highest and second-highest scores per task are highlighted in \greentag{green} and \bluetag{blue}, respectively.
    }  
	\label{tab: main-table-max}
        \resizebox{\textwidth}{!}{%
	\begin{tabular}{ccccc|ccc}
		\toprule
		\textbf{Method} & \textbf{Ant Morphology} & \textbf{D’Kitty Morphology} & \textbf{TF Bind $8$} & \textbf{TF Bind $10$} & \textbf{Mean Score $\uparrow$} & \textbf{Rank Mean $\downarrow$} & \textbf{Rank Median $\downarrow$}\\
		\midrule
		$\mathcal{D}$(\textbf{best}) & $0.565$ & $0.884$ & $0.439$& $0.511$ & $-$ & $-$  & $-$ \\
		\midrule
        % \multicolumn{8}{l}{\textit{Forward Methods}} \\
        % \cmidrule{1-8}
        % Grad-mean & $0.644\pm 0.039$ & $0.907\pm 0.016$ & $0.666\pm 0.011$ & $0.695\pm 0.027$ & $0.728 \pm 0.023$ & $9.5$ & $8.5$ \\
        Grad-mean & $0.709 \pm 0.002$  & $0.920 \pm 0.008$ & $0.843 \pm 0.082$ & \second{0.736 $\pm$ 0.016} & $0.802 \pm 0.027$ & $4.25$ & \second{3.5} \\
        % Grad-EI & $0.626\pm 0.002$ & $0.901\pm 0.045$ & $0.673\pm 0.012$ & $0.689\pm 0.013$ & $0.722 \pm 0.018$ & $10.5$ & $10.5$ \\
        Grad-EI & $0.655 \pm 0.002$ & $0.923 \pm 0.010$ & \second{0.864 $\pm$ 0.091}  & $0.727 \pm 0.024$ & $0.792 \pm 0.032$ & $4.25$ & $4.0$ \\
        COMs & $0.647\pm 0.020$ & $0.934\pm 0.008$ & $0.843\pm 0.046$ & $0.709\pm 0.025$ & $0.783 \pm 0.025$ & $4.5$ & $4.5$ \\
        ICT & $0.555\pm 0.045$ & $0.932\pm 0.037$ & $0.753\pm 0.050$ & $0.585\pm 0.014$ & $0.706 \pm 0.037$ & $10.0$ & $11.5$ \\
        MATCH-OPT & $0.537\pm 0.024$ & $0.925\pm 0.025$ & $0.697\pm 0.008$ & $0.583\pm 0.034$ & $0.686 \pm 0.023$ & $11.75$ & $12.5$ \\
        UniSO-T & $0.636\pm 0.045$ & \second{0.939 $\pm$ 0.007} & $0.836\pm 0.027$ & $0.522\pm 0.017$ & $0.733 \pm 0.024$ & $6.75$ & $6.5$ \\
		\midrule
        % \multicolumn{8}{l}{\textit{Inverse Methods}} \\
        % \cmidrule{1-8}
        CbAS & $0.480\pm 0.019$ & $0.911\pm 0.035$ & $0.721\pm 0.028$ & $0.597\pm 0.005$ & $0.677 \pm 0.022$ & $13.0$ & $13.5$ \\
        ExPT & \second{0.929 $\pm$ 0.049} & \best{0.950\pm 0.041} & $0.810\pm 0.044$ & $0.703\pm 0.022$ & \second{0.848 $\pm$ 0.039} & \second{4.0} & $4.0$ \\
        % \cdashline{1-5}\cdashline{7-8}
        MIN & $0.570\pm 0.003$ & $0.886\pm 0.017$ & $0.764\pm 0.008$ & $0.517\pm 0.030$ & $0.684 \pm 0.015$ & $12.25$ & $12.5$ \\
        % \cdashline{1-5}\cdashline{7-8}
        BONET & $0.632\pm 0.042$ & $0.920\pm 0.040$ & $0.776\pm 0.007$ & $0.492\pm 0.043$ & $0.705 \pm 0.033$ & $10.25$ & $8.5$ \\
        OPRO & $0.517\pm 0.039$ & $0.856\pm 0.046$ & $0.758\pm 0.017$ & $0.500\pm 0.013$ & $0.658 \pm 0.029$ & $14.0$ & $14.5$ \\
        % \cdashline{1-5}\cdashline{7-8}
        GTG & $0.603\pm 0.039$ & $0.917\pm 0.023$ & $0.762\pm 0.016$ & $0.730 \pm 0.026$ & $0.753 \pm 0.026$ & $8.25$ & $9.5$ \\
        DDOM & $0.590\pm 0.026$ & $0.929\pm 0.037$ & $0.739\pm 0.016$ & $0.497\pm 0.002$ & $0.689 \pm 0.020$ & $11.25$ & $12.5$ \\
        % dLLM & $-$ & $-$ & $-$ & $-$ & $-$ & $-$ \\
		% \cdashline{1-5}\cdashline{7-8}
        CMA-ES & $0.592\pm 0.010$ & $0.711\pm 0.045$ & $0.784\pm 0.029$ & $0.658\pm 0.031$ & $0.686 \pm 0.029$ & $10.25$ & $9.0$ \\
        MCTS-transfer & $0.648\pm 0.001$ & $0.910\pm 0.006$ & $0.857 \pm 0.015$ & $0.628\pm 0.043$ & $0.761 \pm 0.016$ & $7.25$ & $6.5$ \\
		\midrule
		\textbf{DiBO}$_{\mathrm{(ours)}}$ & \best{0.932 \pm 0.022} & $0.912 \pm 0.017$ & \best{0.946 \pm 0.043} & \best{0.741 \pm 0.027} & \best{0.883 \pm 0.027} & \best{3.5} & \best{1.0} \\
		\bottomrule
	\end{tabular}
	}
\end{table*}

% \ZP{Ye please help me proofread this part.}

We initialize our model from the pretrained diffusion language model \textsc{LLaDA-8B-Instruct}~\citep{nie2025large}. The training process follows a sequential pipeline: domain adaptation (DA), supervised fine-tuning (SFT), and reinforcement learning (RL). 
For discrete tasks, we employ learning rates of $2\times10^{-5}$ for DA and SFT, and $1\times10^{-6}$ for RL. For continuous tasks, the learning rates are $1\times10^{-5}$ (DA), $2\times10^{-5}$ (SFT), and $1\times10^{-6}$ (RL). 
Each stage begins with a $100$-step linear warmup followed by a constant learning rate schedule.
We use a per-device batch size of $1$, with gradient accumulation steps set to $16$ for discrete tasks and $8$ for continuous tasks. 
\textcolor{black}{
% Across the DA, SFT, and RL stages, we train the model for $2,048$ / $1,024$ / $128$ optimization steps. 
For DA stage, we train the model for $1,024$ and $2,048$ optimization steps for discrete and continuous tasks respectively. For SFT and RL stages, we optimize all tasks with $1,024$ and $128$ steps.
Factoring in gradient accumulation, this equates to processing $16,384$ / $16,384$ / $2,048$ total samples for discrete tasks, and $16,384$ / $8,192$ / $1,024$ samples for continuous tasks.
% Factoring in gradient accumulation, this equates to processing $16,384$ / $16,384$ / $2,048$ and $16,384$ / $8,192$ / $1,024$ total samples for discrete and continuous tasks respectively.
}
% We set the training step to 2048 (DA) and 1024 (SFT), and evaluate on the 128-th step on RL stage for all 4 tasks. \Ye{Move these to appendix or README of our code?}
%
% Training is conducted using the \texttt{PagedAdamW8bit} optimizer as implemented in the \texttt{bitsandbytes} library \cite{dettmers2022llmint88bitmatrixmultiplication, dettmers20228bitoptimizersblockwisequantization} with Bfloat16 precision. 
We adopt the \texttt{PagedAdamW8bit} optimizer as implemented in the
\texttt{bitsandbytes} library, building on block-wise quantized optimizers~\citep{dettmers20228bitoptimizersblockwisequantization},
with Bfloat16 precision.
% Activation checkpointing is disabled, as it is not officially supported in the current LLaDA implementation. \Ye{previous sentence may not be needed} 
All diffusion-related hyperparameters, including the masking schedule and mask ratio, remain consistent with the official implementation by \citet{nie2025large}.

For each task, we construct a fixed offline pool of $n_{\text{pool}}=500$ samples by evenly sub-sampling the label-sorted dataset. 
To remain within the memory constraints of a single NVIDIA H100 ($80$\,GB) GPU, we use $n_{\text{few}}=7$ few-shot examples per prompt. 
\textcolor{black}{
At inference time, for each candidate, we independently resample the corresponding context examples from the offline pool.
We complete the masked response in a single forward pass using greedy token filling without temperature-based stochastic decoding.
Duplicate outputs are discarded, and generation continues until 128 unique valid candidates are obtained.
}
Detailed sampling rules, text rendering formats, and prompt templates are provided in Appendix~\ref{app: offline-data-construction}, \ref{app: prompt-response-construction}, and \ref{app: prompt-templates}. All experiments are performed on a single H100 GPU; for instance, the entire pipeline for the TF Bind $8$ task can be completed within $1.5$ hours. 
% We report the mean and standard deviation across $8$ random seeds for all evaluations.
% We report the mean and standard deviation across $8$ random seeds in all evaluations.
% All results are reported as the mean and standard deviation over 8 random seeds.
All results are reported as mean $\pm$ standard deviation, averaged over $8$ different random seeds.

\subsection{Results and Analysis}
\label{subsec: result_and_analysis}
\begin{table*}[!t]
\centering
\small
\setlength{\tabcolsep}{7pt}
\renewcommand{\arraystretch}{1.15}

\caption{
\textcolor{black}{\textbf{Matched comparison between diffusion and autoregressive (AR) backbones.}
We construct an autoregressive counterpart (LLaMA-3.1-8B-Instruct) by keeping the prompt construction, dataset, delimiter tokens, context selection strategy, and the full post-training pipeline (DA→SFT→RL) unchanged, and replacing only the backbone architecture (diffusion vs. AR). 
Results report the $100$-th percentile normalized oracle score (mean ± std over seeds). 
The last row (“Performance Gain”) shows the absolute improvement of the diffusion backbone over the matched AR backbone at each stage.}
}
\label{tab:ab-ar}

\begin{subtable}{\textwidth}
\centering
\small
\caption{Discrete Tasks}

% \begin{adjustbox}{max width=\textwidth}
    \begin{tabular}{lcccccc}
    % \small
    \toprule
    \multirow{2}{*}{\parbox[c][3.6ex][c]{2.2cm}{\centering \multirow{2}{*}{\centering \vspace{0.2ex} Method}}} 
    & \multicolumn{3}{c}{TF Bind~$8$} & \multicolumn{3}{c}{TF Bind~$10$} \\
    \cmidrule(lr){2-4}\cmidrule(lr){5-7}
    & DA & SFT & RL & DA & SFT & RL \\
    \midrule
    
    Autoregressive
    & $0.803 \pm 0.038$ 
    & $0.875 \pm 0.008$ 
    & $0.915 \pm 0.008$ 
    & $0.623 \pm 0.047$ 
    & $0.633 \pm 0.036$ 
    & $0.682 \pm 0.053$ \\
    
    DiBO$_{\mathrm{(ours)}}$
    & $0.883 \pm 0.032$ 
    & $0.939 \pm 0.031$ 
    & $0.946 \pm 0.043$ 
    & $0.644 \pm 0.039$ 
    & $0.704 \pm 0.056$ 
    & $0.741 \pm 0.027$ \\
    
    \midrule
    
    \rowcolor{lightgreen}
    Performance Gain
    & $0.080\uparrow\uparrow$ 
    & $0.064\uparrow\uparrow$ 
    & $0.031\uparrow$ 
    & $0.021\uparrow$ 
    & $0.071\uparrow\uparrow$ 
    & $0.059\uparrow\uparrow$ \\
    
    \bottomrule
    \end{tabular}
% \end{adjustbox}

\end{subtable}

\vspace{0.8em}

\begin{subtable}{\textwidth}
\small
\centering
\caption{Continuous Tasks}

% \resizebox{\textwidth}{!}{
% \begin{adjustbox}{max width=\textwidth}
    \begin{tabular}{lcccccc}
    % \small
    \toprule
    \multirow{2}{*}{\parbox[c][3.6ex][c]{2.2cm}{\centering \multirow{2}{*}{\centering \vspace{0.2ex} Method}}} 
    & \multicolumn{3}{c}{Ant Morphology} & \multicolumn{3}{c}{D’Kitty Morphology} \\
    \cmidrule(lr){2-4}\cmidrule(lr){5-7}
    & DA & SFT & RL & DA & SFT & RL \\
    \midrule
    
    Autoregressive 
    & $0.857 \pm 0.044$ 
    & $0.894 \pm 0.032$ 
    & $0.930 \pm 0.000$ 
    & $0.888 \pm 0.025$ 
    & $0.905 \pm 0.005$ 
    & $0.912 \pm 0.011$ \\
    
    DiBO$_{\mathrm{(ours)}}$
    & $0.875 \pm 0.026$ 
    & $0.929 \pm 0.017$ 
    & $0.932 \pm 0.022$ 
    & $0.903 \pm 0.014$ 
    & $0.908 \pm 0.009$ 
    & $0.912 \pm 0.017$ \\
    
    \midrule
    
    \rowcolor{lightgreen}
    Performance Gain
    & $0.018\uparrow$ 
    & $0.035\uparrow$ 
    & $0.002\uparrow$ 
    & $0.015\uparrow$ 
    & $0.003\uparrow$ 
    & $0.000$ \\
    
    \bottomrule
    \end{tabular}
% \end{adjustbox}

\end{subtable}

\vspace{-3mm}

\end{table*}

Following previous practice~\citep{yuan2023importance, nguyen2023expt}, Table~\ref{tab: main-table-max} reports the normalized $100$-th percentile oracle scores on four Design-Bench tasks.
Our method, \textbf{DiBO}, achieves the best overall performance, ranking first in terms of mean rank and median rank across all compared methods. 
In addition, we report median (50-th percentile) results in Appendix~\ref{appendix: median-results}, where DiBO also achieves state-of-the-art performance, highlighting the overall effectiveness of our approach.

Across tasks, DiBO consistently outperforms the strongest baseline on three out of four benchmarks---Ant Morphology, TF Bind~$8$,
and TF Bind~$10$.
Notably, on TF Bind~$8$, DiBO surpasses the best baseline by $0.082$ in normalized score, representing a substantial improvement.
On D'Kitty Morphology, DiBO underperforms the best baseline by $0.038$, while remaining competitive with other strong methods.
Overall, these results demonstrate the strong effectiveness of DiBO across diverse tasks and design domains.

% On D'Kitty Morphology, DiBO achieves competitive performance, closely matching other strong baselines while falling short of the
% best-performing method by a modest margin.
% We hypothesize that this behavior is partly due to the relatively high $\mathcal{D}(\text{best})$ value on this task, which reduces
% the available headroom for improvement and makes performance gains more susceptible to saturation effects.
% In such regimes, distinguishing among near-optimal designs becomes increasingly challenging, potentially limiting the observable
% advantage of more expressive generative models.

When compared to \emph{forward methods}, which rely on learned surrogate models and gradient-based optimization,
DiBO consistently demonstrates superior performance.
An explanation is that forward methods are inherently sensitive to surrogate modeling errors and distributional shift:
surrogate models trained on offline data may produce inaccurate gradients when optimizing toward regions that are poorly supported
by the dataset.
In contrast, DiBO directly models the design distribution through generative denoising and avoids explicit reliance on surrogate gradients,
resulting in more robust optimization under limited offline data.

Relative to \emph{inverse methods}, including VAE-, GAN-, autoregressive-, and diffusion-based approaches,
DiBO also achieves strong and often superior performance.
% \ZP{added mcts comparison}. 
% We also consider the inference-only diffusion LLM baseline with MCTS~\citep{yuan2026diffusion} and discuss in Appendix~\ref{appendix:dllm-mcts}.
% While inverse methods benefit from directly modeling high-quality designs, many of them either rely on unidirectional generation
% or operate in continuous design spaces that require task-specific architectural choices.
% DiBO instead performs optimization directly in a unified discrete token space using diffusion language modeling,
% enabling bidirectional context modeling and seamless integration of textual instructions, designs, and labels.
While existing LLM-based optimization methods have shown encouraging performance,
they adopt different modeling assumptions that limit their applicability.
AR LLM approaches such as OPRO~\citep{yang2023large} generate designs sequentially from left to right,
which can make it challenging to capture bidirectional dependencies.
Diffusion-based inverse methods such as DDOM~\citep{krishnamoorthy2023diffusion}
typically operate in continuous design spaces with task-specific architectures,
and do not natively support the integration of textual task descriptions.
DiBO instead performs optimization directly in a unified discrete token space using diffusion language modeling,
enabling bidirectional context modeling and seamless integration of textual instructions, designs, and labels.
This design choice contributes to DiBO’s robust and consistent performance across diverse tasks.
% \ZP{Change the previous sentence.}
% This combination allows DiBO to more effectively exploit the structure of complex design spaces and contributes to its consistent
% advantage across diverse tasks.

\textcolor{black}{
Beyond the main results, Appendix~\ref{app:rna_topk} reports Top-$K$ metrics and three additional RNA design tasks.
% DiBO remains strong across Top-$K$ metrics on the original Design-Bench tasks and comparable on RNA tasks.
Furthermore, we audit whether the pretrained \textsc{LLaDA-8B-Instruct} model already contains task-specific design-label knowledge from public benchmark assets, as detailed in Appendix~\ref{app:pretrain_audit}.
}
Overall, these results demonstrate that DiBO provides a robust and general framework for offline black-box optimization,
achieving strong performance across both discrete sequence and continuous domains while maintaining favorable aggregate rankings across tasks.

\vspace{-2mm}

\paragraph{
\textcolor{black}{Comparison with Inference-Only Diffusion Optimization}
}
\textcolor{black}{
\citet{yuan2026diffusion} also explore diffusion LLMs for black-box optimization, but focus on an inference-only setting that queries a frozen diffusion LLM with MCTS guidance and does not update model parameters.
In contrast, DiBO adapts the diffusion LLM to offline BBO through domain adaptation, supervised fine-tuning, and reinforcement learning, enabling the model to internalize BBO-specific prompt formats, design-label structures, and reward feedback.
Moreover, their method targets an extremely few-shot regime (e.g., $n_{\mathrm{pool}}=10$), whereas our experiments use a small-data offline setting with $n_{\mathrm{pool}}=500$.
We therefore exclude their method from our main baselines and view it as complementary to our training-based framework.
}

% Table~\ref{tab: main-table-max} reports the normalized $100$-th percentile scores on four Design-Bench tasks.
% Our method achieves the best overall performance, ranking first in both mean and median rank across all compared methods.
% On AntMorphology, TFBind8, and TFBind10, we consistently outperform the strongest baseline by approximately $2.5$ points.
% On D'KittyMorphology, we underperforms the best baseline by $2.5$ points, while remaining competitive with other strong methods.
% %this is mainly because the oracle upper bound $\mathcal{D}(\text{best})$ is already high ($0.889$), leaving relatively limited room for improvement.
% Overall, our method shows strong and consistent performance across diverse tasks, achieving the best aggregate ranking.
% %despite a modest degradation on D'KittyMorphology.

\vspace{-1mm}
\subsection{Ablation Studies}
\label{subsec: ablation}
\begin{table*}[t]
\centering
\small
\setlength{\tabcolsep}{10pt}
\renewcommand{\arraystretch}{1.15}
\caption{\textbf{Ablation of the usage of delimiter tokens.}
Plain text (e.g., ``Designs:'' and ``Labels:'') is used to mark boundaries when delimiter tokens are absent.
We report performance after sequential training up to each stage (DA, SFT, and RL) on TF Bind~$8$ and Ant Morphology,
and additionally report the per-stage performance gain brought by using delimiter tokens.
}
\label{tab:ab-delimiters}

\resizebox{\textwidth}{!}{
    \begin{tabular}{lcccccc}
    
    \toprule
    \multirow{2}{*}{\parbox[c][3.6ex][c]{2.2cm}{\centering \multirow{2}{*}{\centering \vspace{0.2ex} Method}}} 
      & \multicolumn{3}{c}{TF Bind~$8$} & \multicolumn{3}{c}{Ant Morphology} \\
    \cmidrule(lr){2-4}\cmidrule(lr){5-7}
      & DA & SFT & RL & DA & SFT & RL \\
    \midrule
    
    w/o delimiter tokens & $0.859 \pm 0.038$ & $0.918 \pm 0.043$ & $0.943 \pm 0.035$ & $0.693 \pm 0.016$ & $0.903 \pm 0.022$ & $0.923 \pm 0.019$ \\
    % w/ delimiters$_{\mathrm{(ours)}}$           & $0.883 \pm 0.032$ & $0.939 \pm 0.031$ & \best{0.965 \pm 0.038} & $0.875 \pm 0.265$ & $0.936 \pm 0.019$ & \best{0.945 \pm 0.016} \\
    w/ delimiter tokens$_{\mathrm{(ours)}}$           & $0.883 \pm 0.032$ & $0.939 \pm 0.031$ & $0.946 \pm 0.043$  & $0.875 \pm 0.026$ & $0.929 \pm 0.017$ & $0.932 \pm 0.022$ \\
    \midrule
    \rowcolor{lightgreen}
    Performance Gain & $0.024\uparrow$ & $0.021\uparrow$ & $0.003\uparrow$ & $0.182\uparrow\uparrow$ & $0.026\uparrow$ & $0.009\uparrow$ \\
    \bottomrule
    \end{tabular}
}
\vspace{-2mm}
\end{table*}
% \begin{table}[t]
% \centering
% \small
% \setlength{\tabcolsep}{10pt}
% \renewcommand{\arraystretch}{1.15}
% \caption{
% Ablation of training stages.
% We evaluate variants obtained by removing one or more stages from the three-stage pipeline
% (domain adaptation, supervised fine-tuning, and reinforcement learning).
% ``--'' indicates that no valid output is produced due to formatting issues.
% }
% \label{tab:ab-stages}

% \resizebox{0.48\textwidth}{!}{
%     \begin{tabular}{lcc}
%     \toprule
%     Setting & TF Bind~8 & Ant Morphology  \\
%     \midrule
    
%     % \multicolumn{3}{l}{\textit{Context length}} \\
    
%     DA-only & $0.882 \pm 0.031$ & $0.875 \pm 0.265$ \\
%     SFT-only & $0.938 \pm 0.034$ & $0.898 \pm 0.027$ \\
%     RL-only & $0.437 \pm 0.001$ & $-$ \\
%     \cmidrule{1-3}
%     DA+SFT & $0.939 \pm 0.030$ & $0.936 \pm 0.019$ \\
%     DA+RL  & $0.928 \pm 0.052$ & $0.858 \pm 0.032$ \\
%     SFT+RL  & $0.951 \pm 0.030$ & $0.913 \pm 0.024$ \\
%     \cmidrule{1-3}
%     DA+SFT+RL$_{\mathrm{(ours)}}$  & $0.965 \pm 0.038$ & $0.945 \pm 0.016$ \\
%     \bottomrule

%     \end{tabular}
% }

% \end{table}

\begin{table}[t]
\centering
% \small
\setlength{\tabcolsep}{8pt}
\renewcommand{\arraystretch}{1.2}
\caption{
% Ablation of training stages.
% We evaluate variants obtained by removing one or more stages from the three-stage pipeline
% (domain adaptation, supervised fine-tuning, and reinforcement learning), and report performance across two tasks.
% All performance drops ($\Delta$) are measured relative to the full pipeline (\textit{DA+SFT+RL}) in the last line.
% ``\xmark'' indicates that no valid output is produced due to formatting issues.
\textbf{Ablation of training stages.} We remove certain training stages from the three-stage pipeline.
Performance drop $\Delta$ is measured relative to the full pipeline (DA+SFT+RL). ``\xmark'' indicates no valid output is produced due to formatting issues.
}
\label{tab:ab-stages}
\resizebox{0.48\textwidth}{!}{
    \begin{tabular}{ccc cc cc}
    \toprule
    \multicolumn{3}{c}{Stages} & \multicolumn{2}{c}{TF Bind $8$} & \multicolumn{2}{c}{Ant Morphology} \\
    \cmidrule(lr){1-3} \cmidrule(lr){4-5} \cmidrule(lr){6-7}
    DA & SFT & RL & Score & $\Delta$ & Score & $\Delta$ \\
    \midrule
    \cmark &       &       & $0.883 \pm 0.032$ & $-0.063\downarrow$ & $0.875 \pm 0.026$ & $-0.057\downarrow$ \\
           & \cmark &       & $0.938 \pm 0.034$ & $-0.008\downarrow$ & $0.898 \pm 0.027$ & $-0.034\downarrow$ \\
           &       & \cmark & $0.437 \pm 0.001$ & $-0.509\downarrow\downarrow$ & \xmark & -- \\
    \midrule
    \cmark & \cmark &       & $0.939 \pm 0.031$ & $-0.007\downarrow$ & $0.929 \pm 0.017$ & $-0.003\downarrow$ \\
    \cmark &       & \cmark & $0.928 \pm 0.052$ & $-0.018\downarrow$ & $0.858 \pm 0.032$ & $-0.074\downarrow$ \\
           & \cmark & \cmark & $0.941 \pm 0.030$ & $-0.005\downarrow$ & $0.913 \pm 0.024$ & $-0.019\downarrow$ \\
    \midrule
    \rowcolor[HTML]{E8F5E9}
    \cmark & \cmark & \cmark & $\mathbf{0.946 \pm 0.043}$ & -- & $\mathbf{0.932 \pm 0.022}$ & -- \\
    \bottomrule
    \end{tabular}
}
% \vspace{-4mm}
\end{table}

% \ZP{Ye please help me check the wording here. Especially notice if I'm over-claiming something dangerous.}

We ablate the key components of our training pipeline to assess their individual contributions,
% All ablations are conducted under the same evaluation protocol as the main experiments.
% All ablations are conducted under the same evaluation protocol as the main experiments.
conducted under the same evaluation protocol as the main experiments.

% We ablate the key components of our training pipeline to assess their individual contributions under the same evaluation protocol as the main experiments for comparison.
% We ablate the key components of our training pipeline to assess their individual contributions.

% \vspace{-3mm}

\paragraph{
\textcolor{black}{Matched Diffusion vs.\ Autoregressive Modeling}
}

\textcolor{black}{
One of our central motivations is that many design problems exhibit bidirectional dependencies, where each design variable can be influenced by both its prefix and suffix.
To this end, we conduct a matched comparison between diffusion and autoregressive (AR) backbones, by replacing \textsc{LLaDA-8B-Instruct} with \textsc{LLaMA-3.1-8B-Instruct} while keeping all other settings unchanged.
}

\textcolor{black}{
As shown in Table~\ref{tab:ab-ar}, the diffusion backbone outperforms or matches the corresponding AR backbone across all tasks and training stages.
The gains are especially pronounced on the discrete DNA sequence tasks, where diffusion improves over AR by up to $0.080$ on TF Bind~$8$ and $0.071$ on TF Bind~$10$.
This supports our hypothesis that bidirectional refinement is particularly beneficial for structured design problems, where functional dependencies are often non-sequential.
The gains are smaller but still positive on the continuous tasks, suggesting that the advantage of diffusion modeling also extends beyond discrete sequence generation.
}

% \vspace{-4mm}

\paragraph{Delimiter Tokens}
We remove the design and label delimiter tokens  (1)\texttt{|design-start|}/\texttt{|design-end|} and 
(2)\texttt{|label-start|}/\texttt{|label-end|} and train the model using plain text inputs (e.g., ``Designs: '' and ``Labels: '').
Performance is evaluated after sequential training up to each stage as shown in Table~\ref{tab:ab-delimiters}.

Removing these delimiters leads to consistent performance degradation across all three training stages (DA, SFT, and RL) on both tasks.
The performance drop is particularly pronounced during the domain adaptation stage on Ant Morphology, which further underscores the importance of delimiter tokens in enabling effective domain adaptation, especially for complex continuous spaces.

Overall, these results demonstrate that explicit delimiter tokens play a critical role throughout the entire training pipeline,
by providing a clear separation between heterogeneous input components and facilitating effective learning across different stages.

\begin{table}[t]
\centering
\small
\setlength{\tabcolsep}{10pt}
\renewcommand{\arraystretch}{1.15}
\caption{
\textbf{Ablation on similarity-conditioned context construction.}
We compare selecting prompt examples based on design similarity against removing this locality constraint and using randomly selected context examples.
}
\label{tab:ab-neighbor}

\resizebox{0.48\textwidth}{!}{
    \begin{tabular}{lcc}
    \toprule
    Setting & TF Bind~$8$ & Ant Morphology  \\
    \midrule
    \cmidrule{1-3}
    Random Context & $0.730 \pm 0.001$ & $0.906 \pm 0.008$ \\
    Similar Context$_{\mathrm{(ours)}}$ & $0.946 \pm 0.043$  & $0.932 \pm 0.022$ \\
    \midrule
    \rowcolor{lightgreen}
    Performance Gain & $0.216\uparrow\uparrow$ & $0.026\uparrow$ \\
    \bottomrule
    \end{tabular}
}
\vspace{-4mm}
\end{table}

% \vspace{-2mm}

\paragraph{Training Stages}
We evaluate variants that remove one stage from the three-stage pipeline (\textit{DA+SFT+RL}), resulting in \textit{DA+SFT}, \textit{SFT+RL}, and \textit{DA+RL}.
As shown in Table~\ref{tab:ab-stages}, across both tasks, removing any stage leads to a clear performance drop compared to the full pipeline, highlighting the importance of all three stages.

% Among 2-stages variants, we observe that \textit{DA+RL} exhibits the largest degradation in performance, suggesting the reinforcement signal alone is insufficient to fully exploit fine-grained designs without an intermediate supervised fine-tuning stage.
% The \textit{SFT+RL} variant also underperforms the full pipeline, indicating that learning without prior domain adaptation is suboptimal, as the model’s understanding of the heterogeneous input components remains limited.

Among the two-stage variants, \textit{DA+RL} exhibits the largest degradation in performance,
% (TF Bind 8: $\Delta=-0.037$; Ant: $\Delta=-0.087$),
indicating that reinforcement learning without an intermediate supervised stage is insufficient to reliably exploit fine-grained prompt--response mappings, especially in the continuous domain.
Similarly, \textit{SFT+RL} underperforms the full pipeline,
% (TF Bind 8: $\Delta=-0.014$; Ant: $\Delta=-0.032$)
suggesting that skipping domain adaptation leads to suboptimal learning when handling heterogeneous prompt components.
% \textit{DA+SFT} also exhibits a consistent gap,
% % (TF Bind 8: $\Delta=-0.026$; Ant: $\Delta=-0.009$),
% highlighting our full pipeline benefits from incorporating reinforcement learning as a fine-grained optimization stage on top of a well-aligned representation.
\textit{DA+SFT} also exhibits a consistent gap, highlighting that our full pipeline benefits from incorporating reinforcement learning as a fine-grained optimization stage on top of a well-aligned representation space.

% All single-stage variants perform substantially worse than the integrated approach.
% Notably, the RL-only model fails to produce valid outputs on Ant Morphology,
% demonstrating that domain adaptation is essential for learning the correct input-output structure in complex design tasks.
% Comparing \textit{DA-only} and \textit{DA+SFT}, we observe that the gain from domain adaptation alone is more pronounced on Ant Morphology, where stronger adaptation is required to learn valid design representations in a complex continuous space.
% Overall, these results confirm that the three training stages are complementary and jointly necessary to achieve strong and consistent performance across tasks.

All single-stage variants perform substantially worse than the integrated approach.
In particular, \textit{RL-only} fails to produce valid outputs on Ant Morphology, highlighting that RL without prior alignment struggles to correctly learn the heterogeneous components in complex design tasks.
% Finally, comparing \textit{DA-only} to \textit{DA+SFT}, adding SFT after DA yields a similar boost on both tasks (TF Bind 8: $+0.057$; Ant: $+0.061$), suggesting that supervised post-training is a generally effective bridge from domain adaptation to optimization-oriented learning.
% Overall, these results confirm that the three training stages are complementary and jointly necessary to achieve strong and consistent performance across tasks.
Overall, these results confirm that the three training stages are complementary and jointly necessary to achieve strong and consistent performance across tasks.
In particular, these results emphasize the importance of a well-structured training pipeline, where different stages play complementary roles rather than acting in isolation.

\begin{table*}[ht]
\centering
\small
\setlength{\tabcolsep}{8pt}
\renewcommand{\arraystretch}{1.15}

\caption{
\textcolor{black}{
\textbf{Comparison between evenly spaced sub-sampling and random sub-sampling of the offline dataset.}
We report the $100$th-percentile normalized oracle score on four tasks, as well as the performance difference between the two strategies.
}
}

\label{tab:ab-randomsubsampling}

\begin{tabular}{lcccc}

\toprule
Method & Ant Morphology & D’Kitty Morphology & TF Bind~$8$ & TF Bind~$10$ \\
\midrule

Random sub-sampling & $0.936 \pm 0.019$  & $0.909 \pm 0.008$ & $0.960 \pm 0.027$ & $0.743 \pm 0.012$ \\
Evenly sub-sampling$_{\mathrm{(ours)}}$ & $0.932 \pm 0.022$ & $0.912 \pm 0.017$ & $0.946 \pm 0.043$   & $0.741 \pm 0.027$ \\

\midrule
Difference & $-0.004$ & $+0.003$ & $-0.014$  & $-0.002$ \\

\bottomrule
\end{tabular}

\end{table*}

\begin{figure*}[ht]
\centering
\begin{minipage}[t]{0.24\textwidth}
  \centering
  \includegraphics[width=\linewidth]{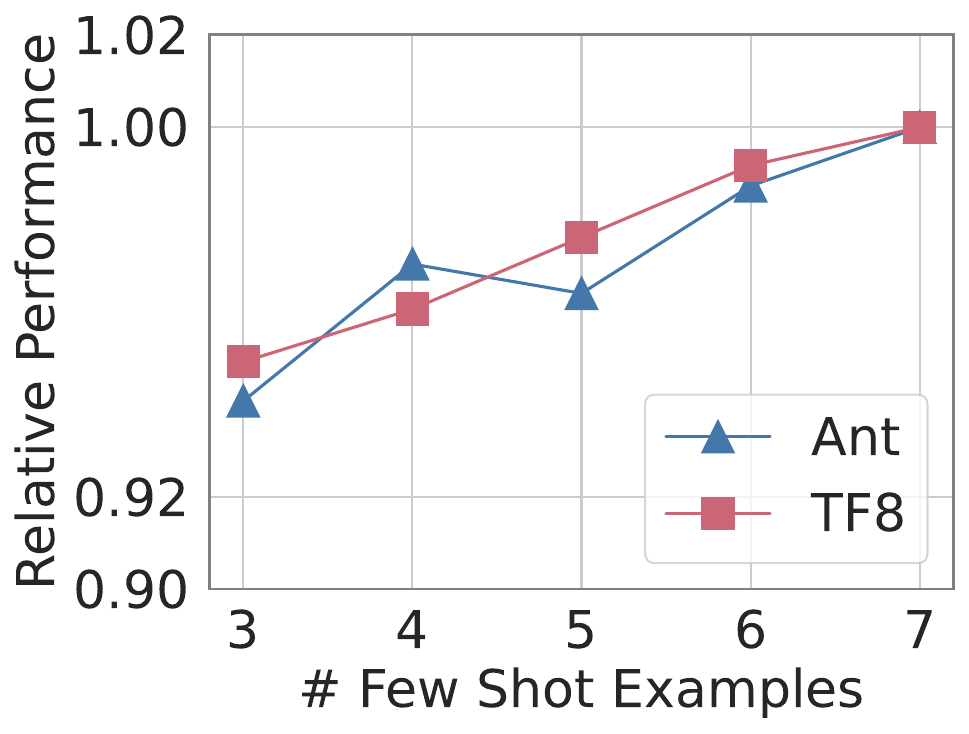}
  \vspace{-2mm}
  {\small (a) Context length}
\end{minipage}\hfill
\begin{minipage}[t]{0.24\textwidth}
  \centering
  \includegraphics[width=\linewidth]{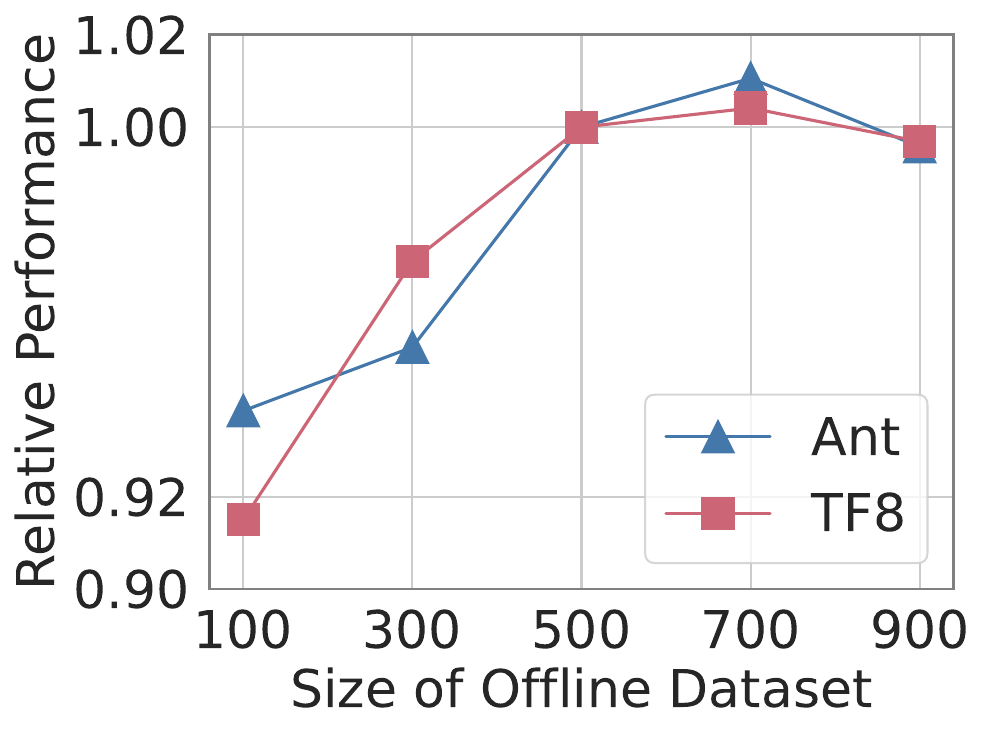}
  \vspace{-2mm}
  {\small (b) Size of the offline dataset}
\end{minipage}\hfill
\begin{minipage}[t]{0.24\textwidth}
  \centering
  \includegraphics[width=\linewidth]{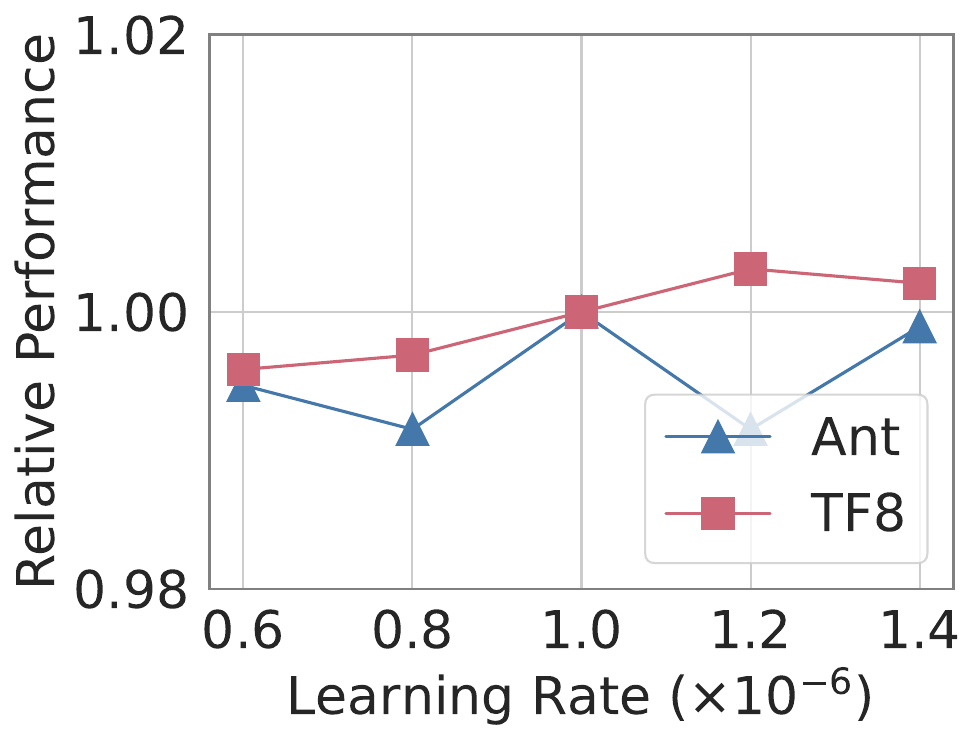}
  \vspace{-2mm}
  {\small (c) Learning rate}
\end{minipage}\hfill
\begin{minipage}[t]{0.24\textwidth}
  \centering
  \includegraphics[width=\linewidth]{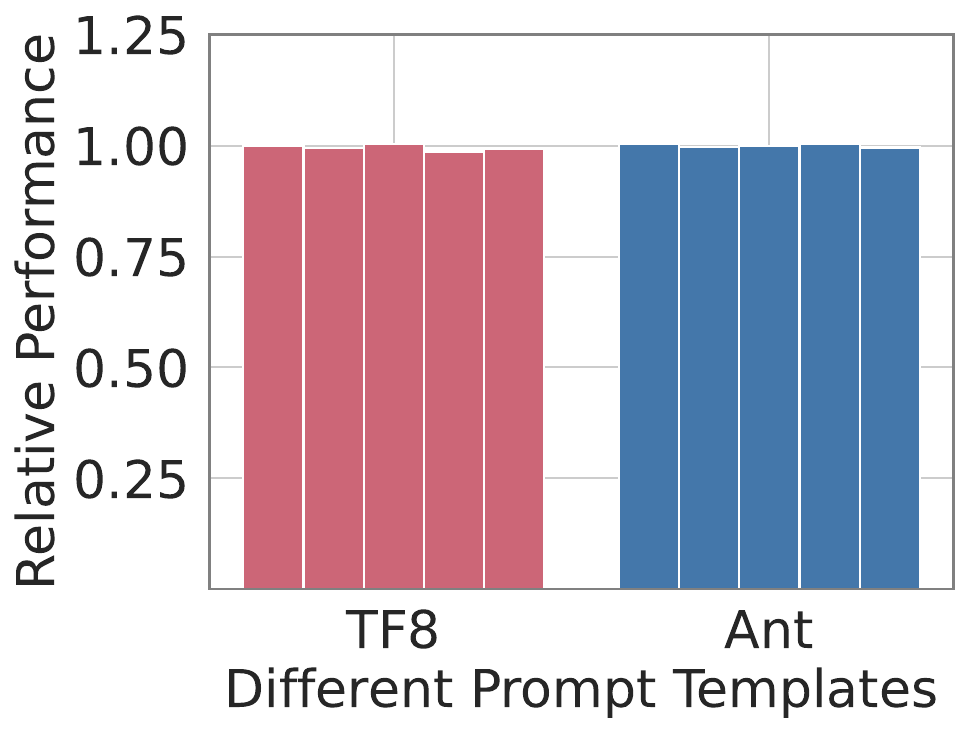}
  \vspace{-2mm}
  {\small (d) Prompt templates}
\end{minipage}
\vspace{1mm}
\caption{
    \textbf{Hyperparameter sensitivity} %on TF Bind~8 and Ant Morphology 
    at the RL stage.
    % We evaluate the robustness of our method on TF Bind8 and Ant Morphology with respect to key hyperparameters: 
    % (a) the number of few-shot examples in the prompt (context length), (b) size of the offline dataset, (c) learning rate used in RL stage, and (d) variations of prompt templates. 
    Results are reported as relative performance normalized by our default setting. 
    % Overall, performance improves with increased context length and larger offline pools, while remaining stable across a wide range of learning rates and prompt formulations, indicating robust RL optimization behavior.
}
\label{fig:hyper-4panel}
\vspace{-4mm}
\end{figure*}

\vspace{-2mm}

\paragraph{Similarity-Conditioned Context Construction}
% \input{tables/ab-neighbors}
% \ZP{Should change.}
We study the effect of conditioning the target response to remain within a reasonable distribution of the prompt examples.
% locally consistent with the target response, 
%
As shown in Table~\ref{tab:ab-neighbor}, this similarity-conditioned context
is critical for performance.
Removing this constraint leads to a substantial degradation on both tasks,
with a particularly severe drop on TF Bind~$8$.
These results indicate that diffusion LLMs benefit from prompt contexts that
preserve local semantic consistency, which facilitates effective reasoning
over the underlying design landscape during generation.

\vspace{-2mm}

\paragraph{Evenly Spaced Sub-Sampling}
\textcolor{black}{
% We further examine whether constructing the offline pool by evenly sub-sampling the label-sorted dataset affects the final performance.
% Our motivation for this deterministic strategy is to ensure reproducibility and maintain coverage across the full label spectrum.
% As shown in Table~\ref{tab:ab-randomsubsampling}, evenly spaced sub-sampling and random sub-sampling lead to very similar performance across all four tasks, with differences around $0.01$ normalized score, suggesting our results not sensitive to the particular offline-pool construction strategy.
We examine the effect of evenly sub-sampling the label-sorted dataset, which provides reproducible coverage of the label spectrum.
As shown in Table~\ref{tab:ab-randomsubsampling}, evenly spaced and random sub-sampling achieve similar performance across all four tasks, with differences around $0.01$ normalized score.
Thus, our results are not sensitive to this offline-pool sampling strategy.
}

% \paragraph{Audit}
% \textcolor{red}{
% audit pretrained model. May not be able to appear in main text due to length constaint. lets see.
% }

\vspace{-1mm}

\subsection{Hyperparameter Sensitivity}
\label{subsec: hyper}

We evaluate the sensitivity of our method to several critical hyperparameters. Unless otherwise specified, all experiments are conducted using the primary configuration detailed in Section~\ref{subsec: implementation_detail}.

% \paragraph{Size of the offline dataset.}
% We vary the offline pool size $n_{\text{pool}}$ within the set $\{100, 300, 500, 700, 900\}$. As shown in Table~\ref{tab:ab-fewshot-npool}, performance improves consistently as the pool size increases, with the highest performance achieved at $n_{\text{pool}}=900$. This trend suggests that larger offline pools provide more comprehensive coverage of the design landscape, although we observe diminishing marginal gains as the pool size exceeds $700$.

\vspace{-1.5mm}

\paragraph{Context Length}
% We investigate the impact of the number of few-shot examples provided in the prompt, varying the count from $3$ to $7$ (see Figure~\ref{fig:hyper-4panel} (a)). Performance demonstrates a steady upward trend as more examples are included. However, due to computational memory constraints, we capped the context length at $7$ examples. Consequently, we utilize $7$ few-shot examples for all main experiments to maximize in-context information while remaining within hardware limits.
We investigate the impact of the number of few-shot examples in the prompt, varying the count from $3$ to $7$ (see Figure~\ref{fig:hyper-4panel}(a)). Performance demonstrates a steady upward trend as more examples are included. However, due to GPU memory constraints, we capped the context length at $7$ examples. Consequently, we utilize $7$ few-shot examples for all main experiments to maximize in-context information while remaining within hardware limits.

\vspace{-1.5mm}

\paragraph{Size of Offline Dataset}
We vary the offline pool size $n_{\text{pool}}$ within $\{100, 300, 500, 700, 900\}$, where the default size is $500$.
As shown in Figure~\ref{fig:hyper-4panel}(b), performance is worst when only $100$ samples are used, and improves consistently as the pool size increases on both tasks. A slight drop when $n_{\text{pool}}$ reaches $900$ is observed.

\vspace{-1.5mm}

\paragraph{Learning Rate of Reinforcement Learning}
To assess the stability of the Reinforcement Learning (RL) phase, we vary the learning rate among $\{0.6, 0.8, 1.0, 1.2, 1.4\} \times 10^{-6}$ and report the results in Figure~\ref{fig:hyper-4panel}(c). We observe that the model's performance remains largely invariant across this range, suggesting that our RL training objective is robust and relatively insensitive to specific learning rate tuning.

\vspace{-6mm}

% \paragraph{Prompt template variation.}
% % \ZP{I will illustrate the results in figures directly for this part.}
% To evaluate robustness to linguistic variations, we generated five distinct sets of prompt templates using GPT-5.2 and re-trained the model on the TF Bind 8 and Ant Morphology tasks. As illustrated in Figure~\ref{fig:hyper-4panel} (d), performance across these variations remained stable with negligible fluctuations. These results indicate that our method primarily relies on the underlying optimization structure and data patterns rather than specific semantic choices or ``prompt hacking''.

\paragraph{
\textcolor{black}{Prompt Template Variation}
}
\textcolor{black}{
We evaluate sensitivity to natural prompt paraphrases by using five sets of prompt templates generated by ChatGPT-5.2 and re-training the model on TF Bind~$8$ and Ant Morphology.
These templates preserve the same task semantics, delimiter tokens, and output format, but vary the surface-level instruction wording.
As shown in Figure~\ref{fig:hyper-4panel}(d), performance remains stable across these paraphrastic variants.
% This suggests that DiBO is not dependent on a specific prompt phrasing, although we do not claim robustness to adversarial or poorly specified prompts.
These results indicate that DiBO is stable under natural paraphrases of the task instruction, as long as the task semantics and structured output format are preserved.
More details on the prompt examples we use are provided in Appendix~\ref{app: prompt-templates}. 
}

\vspace{-2mm}

\section{Conclusion}

We present DiBO, a framework for adapting diffusion language models to black-box optimization. To bridge the semantic gap between natural language and offline design-label data, we introduce explicit delimiter tokens and a joint prompt--response reconstruction loss for domain adaptation. We further combine supervised fine-tuning and reinforcement learning in a two-stage post-training pipeline, enabling the model to synthesize high-label designs from complex, heterogeneous inputs. Experiments across diverse discrete and continuous Design-Bench tasks show strong and consistent performance, underscoring the promise of diffusion language models as a robust, 
general-purpose paradigm for offline optimization in high-dimensional design spaces.

\section*{Impact Statement}
% \ZP{Need to check.}

This paper presents a diffusion-based language model framework for offline black-box optimization, aiming to accelerate discovery in scientific and engineering domains such as molecular sequence design, robotics, and materials science. By reducing reliance on resource-intensive physical experiments and avoiding online trial-and-error, our approach offers significant advantages in cost-efficiency and safety for sensitive applications.

However, as a general-purpose generative optimization method, it carries potential risks if applied without appropriate safeguards. The model may inherit or amplify biases present in the offline training datasets. Furthermore, while intended to assist scientific discovery, we acknowledge the risk of misuse in generating harmful designs (e.g., in biochemical contexts). We emphasize that this tool is designed to augment expert judgment, not replace it, and its deployment requires responsible oversight.

% \section{Some Notes}
% https://github.com/ML-GSAI/LLaDA-1.5
% Use the two special tokens in SFT/RL process.
% this paper summarizes https://arxiv.org/abs/2508.10875 how to do SFT/RL.

% MMaDA: Multimodal Large Diffusion Language Models
% We could the RL prob estimation in this paper.

\bibliography{main}
\bibliographystyle{icml2026}

\newpage
\appendix
\onecolumn
\section{Appendix}

\subsection{Offline Data Construction}
\label{app: offline-data-construction}
For each Design-Bench task, we load the original designs $\{x^{\text{raw}}_i\}$ and their corresponding scalar labels $\{y_i\}$.
Here, $x^{\text{raw}}_i$ denotes the task-specific raw representation of a design (e.g., discrete sequences for TF Bind~$8$~/~$10$ tasks or continuous vectors for robot design tasks).
All samples are sorted by label value, and an evenly spaced subset of size $n_{\text{pool}}$ is selected to form the offline dataset.
Unless otherwise stated, the offline pool size $n_{\text{pool}}$ is set to $500$ in all main experiments.
This pool remains fixed and is used across all training stages of our method as well as for all comparison methods.

\subsection{Prompt-Response Construction}
\label{app: prompt-response-construction}

\paragraph{Partitioning Strategy}
\label{app: partitioning}
We partition the offline data into two disjoint subsets, $D_1$ and $D_2$, with a fixed ratio of $0.8:0.2$.
$D_1$ corresponds to the sampling pool for prompt context (designs with lower scores), while $D_2$ corresponds to the sampling pool for response targets (designs with higher scores).
We strictly separate $D_1$ and $D_2$ to prevent data leakage; specifically, we ensure that no design used as a target response appears in the few-shot prompt context.

\paragraph{Design Similarity}
\label{app: design-similarity}
Forming prompt-response pairs by randomly sampling from $D_1$ and $D_2$ is often suboptimal. If the designs in the prompt context are structurally dissimilar to the target design in the response, the model faces a significant challenge in learning the underlying local optimization landscape. To address this, we construct prompt-response pairs by selecting designs from the prompt context that are most similar to the target response. Furthermore, for the RL phase, we balance the offline dataset to include an approximately equal distribution of positive and negative reward signals (i.e., improvements vs. degradations relative to the anchor).

We quantify the similarity between designs using a normalized representation, $x^{\text{norm}}$. For sequence-based tasks (e.g., TF Bind~$8$~/~$10$), $x^{\text{norm}}$ is obtained by applying \texttt{map\_to\_logits()} to the discrete sequence, followed by feature-wise normalization. For continuous robot tasks, we directly utilize the normalized continuous features provided by Design-Bench. The similarity between two designs $i$ and $j$ is defined using an RBF kernel:
\[
\operatorname{sim}(i,j) = \exp\!\left(-\frac{\|x^{\text{norm}}_i - x^{\text{norm}}_j\|_2^2}{2\sigma^2}\right),
\]
where the kernel bandwidth $\sigma$ is the median of all pairwise Euclidean distances computed across the entire offline pool.

\subsection{Prompt Templates}
\label{app: prompt-templates}

We use a total of $10$ natural-language prompt templates to introduce linguistic diversity:
$8$ templates are used for training, and $2$ templates are held out for validation.
All templates preserve the same optimization objective, delimiter tokens, design-label structure, and output format, and differ only in surface-level natural-language wording. The paraphrases are generated by ChatGPT-5.2.

% \begin{table}[ht]
% \centering
% \small
% \setlength{\tabcolsep}{4pt}
% \caption{
% \textcolor{red}{
% Representative prompt paraphrases for TF Bind 8. All variants preserve the same task semantics, delimiter tokens, and output format, while varying only the surface-level instruction wording.}
% }
% \label{tab:prompt_variants}
% \begin{tabular}{@{}c p{0.85\linewidth}@{}}
% \toprule
% Template & Instruction wording \\
% \midrule
% T1 & You are a helpful optimization assistant that will help us generate a new length-8 optimal DNA sequence with maximum binding affinity with a particular transcription factor SIX6 REF R1. \\
% T2 & The goal is to design an 8-length DNA sequence with maximum binding affinity to the transcription factor SIX6 REF R1." \\
% T3 & Optimization Task: Find a DNA sequence (length = 8) that binds more strongly to SIX6 REF R1. Training data (sequence, affinity) is as follows. \\
% T4 & You are assisting a molecular biologist in optimizing DNA binding sites. \\
% T5 & Below is a reference dataset of DNA sequences ranked by binding affinity to the transcription factor SIX6 REF R1. \\
% \bottomrule
% \end{tabular}
% \end{table}

\begin{table}[ht]
\centering
\small
\setlength{\tabcolsep}{3pt}
\caption{
\textcolor{black}{
\textbf{Representative prompt paraphrases for TF Bind $8$.} 
All variants preserve the same task semantics, delimiter tokens, and output format, while varying only the surface-level instruction wording.}
}
\label{tab:prompt_variants}
\begin{tabularx}{\linewidth}{@{}>{\centering\arraybackslash}m{0.10\linewidth}>{\raggedright\arraybackslash}X@{}}
\toprule
Template & Instruction wording \\
\midrule
T1 & You are a helpful optimization assistant that will help us generate a new length-$8$ optimal DNA sequence with maximum binding affinity with a particular transcription factor \texttt{SIX6\_REF\_R1}. \\
T2 & The goal is to design an $8$-length DNA sequence with maximum binding affinity to the transcription factor \texttt{SIX6\_REF\_R1}. \\
T3 & Optimization Task: Find a DNA sequence (length = $8$) that binds more strongly to \texttt{SIX6\_REF\_R1}. Training data (sequence, affinity) is as follows. \\
T4 & You are assisting a molecular biologist in optimizing DNA binding sites. \\
T5 & Below is a reference dataset of DNA sequences ranked by binding affinity to the transcription factor \texttt{SIX6\_REF\_R1}. \\
\bottomrule
\end{tabularx}
\end{table}

% \begin{table}[ht]
% \centering
% \small
% \setlength{\tabcolsep}{4pt}
% \caption{
% \textcolor{red}{
% Representative prompt paraphrases for TF Bind 8. All variants preserve the same task semantics, delimiter tokens, and output format, while varying only the surface-level instruction wording.}
% }
% \label{tab:prompt_variants}

% \begingroup
% \renewcommand{\tabularxcolumn}[1]{>{\raggedright\arraybackslash}m{#1}}
% \begin{tabularx}{\linewidth}{@{}>{\centering\arraybackslash}m{0.10\linewidth}X@{}}
% \toprule
% Template & Instruction wording \\
% \midrule
% T1 & You are a helpful optimization assistant that will help us generate a new length-8 optimal DNA sequence with maximum binding affinity with a particular transcription factor SIX6 REF R1. \\
% T2 & The goal is to design an 8-length DNA sequence with maximum binding affinity to the transcription factor SIX6 REF R1. \\
% T3 & Optimization Task: Find a DNA sequence (length = 8) that binds more strongly to SIX6 REF R1. Training data (sequence, affinity) is as follows. \\
% T4 & You are assisting a molecular biologist in optimizing DNA binding sites. \\
% T5 & Below is a reference dataset of DNA sequences ranked by binding affinity to the transcription factor SIX6 REF R1. \\
% \bottomrule
% \end{tabularx}
% \endgroup
% \end{table}

\paragraph{Paraphrased Template Examples}
\textcolor{black}{
Table~\ref{tab:prompt_variants} shows representative prompt-instruction variants for TF Bind $8$.
All variants describe the same optimization goal (i.e., designing a length-8 DNA sequence with improved binding affinity to the same transcription factor), but use different surface-level wording.
For example, some variants state the task as a direct goal, while others introduce it as a reference dataset or an optimization problem.
After this instruction part, all variants use the same structured design-label context with \texttt{|design-start|}, \texttt{|design-end|}, \texttt{|label-start|}, and \texttt{|label-end|}, and require the response to follow the same delimiter-based design format.
}

% \paragraph{Example templates.}
% Below we show one example prompt for each task.

\paragraph{Example Prompt Instantiations}
\textcolor{black}{
We further provide one full prompt instantiation for each task as better reference.
These examples illustrate how the task description, few-shot design-label context, delimiter tokens, and response format are combined in the actual inputs.
For continuous tasks, vector entries are abbreviated with ellipses for readability, while the actual inputs contain the full design vectors.
}

% (Recommended) Put the font size in the tcolorbox options (not as \small inside),
% and use tcblisting/Verbatim-like blocks for the long code-like lines.

% If you prefer keeping \texttt (instead of \ttfamily blocks), at least fix your
% missing \texttt around the closing delimiters and avoid \small in-body:

\begin{tcolorbox}[title=TF Bind $8$ (Discrete sequence), breakable]
You are a helpful optimization assistant tasked with generating a length-$8$ DNA
sequence that maximizes binding affinity with a specific transcription factor
(\texttt{SIX6\_REF\_R1}).

You are given several existing DNA sequences along with their corresponding
binding affinities:

\texttt{|design-start|['A','C','G','T','A','C','G','T']|design-end|}
\texttt{|label-start|+000.018|label-end|}

\texttt{|design-start|['T','C','G','C','T','A','G','G']|design-end|}
\texttt{|label-start|+000.027|label-end|}

Please propose a new DNA sequence that is different from the existing sequences
and has a higher binding affinity.

Response:
\texttt{|design-start|['G','C','G','T','A','C','G','T']|design-end|}
\end{tcolorbox}

\begin{tcolorbox}[title=TF Bind $10$ (Discrete sequence), breakable]
You are a helpful optimization assistant tasked with generating a length-10 DNA
sequence that optimizes binding to a specific transcription factor (\texttt{Pho4}).

Goal: maximize $-{\mathrm{ddG}}$ (larger values indicate stronger binding).

You are given several existing DNA sequences along with their corresponding $-{\mathrm{ddG}}$ values:

\texttt{|design-start|['T','C','C','A','C','G','A','A','G','A']|design-end|}
\texttt{|label-start|-001.860|label-end|}

\texttt{|design-start|['G','C','T','T','G','G','A','A','C','A']|design-end|}
\texttt{|label-start|+000.010|label-end|}

Please propose a new DNA sequence that is different from the existing sequences
and achieves a higher $-{\mathrm{ddG}}$ value than all sequences shown above.
The DNA sequence should use only \texttt{A}, \texttt{C}, \texttt{G}, and \texttt{T},
and it should differ from each existing sequence in at least one position.

Response:
\texttt{|design-start|['A','C','G','T','A','C','G','T','A','C']|design-end|}
\end{tcolorbox}

\begin{tcolorbox}[title=AntMorphology (Continuous control), breakable]
You are a helpful optimization assistant for Ant robot morphology design in OpenAI Gym.
The goal is to maximize the running performance of a simulated quadruped robot (Ant).
Each morphology is represented by $60$ continuous parameters ($4$ legs $\times$ $15$ parameters per leg).

For each leg, parameters are ordered as follows:
$x$ (hip x), $y$ (hip y), $z$ (hip z),
$a$ (hip angle), $b$ (thigh angle), $c$ (ankle angle),
hip center, hip range,
thigh center, thigh range,
ankle center, ankle range,
hip size, thigh size, ankle size.

Existing designs and their performance scores:

\texttt{|design-start|[+001.000, +010.000, \ldots, -002.345, +000.091]|design-end|}
\texttt{|label-start|+000.742|label-end|}

\texttt{|design-start|[+002.500, +008.750, \ldots, -001.120, +000.305]|design-end|}
\texttt{|label-start|+000.768|label-end|}

Please propose one new morphology design that is different from the above designs
and is expected to achieve a higher performance score.
All parameters must be floating-point numbers rounded to three decimal places.

Response:

\texttt{|design-start|[+002.000, +009.500, \ldots, -001.800, +000.250]|design-end|}
\end{tcolorbox}

\begin{tcolorbox}[title=D'KittyMorphology (Continuous control), breakable]
You are a helpful optimization assistant for D'Kitty robot morphology design in simulation.

Goal: maximize the performance score (better locomotion/navigation to a fixed target; higher is better).

Each design is $56$ continuous parameters ($4$ legs $\times$ $14$ parameters per leg).

Per-leg parameters (in order):
$x$ (hip x), $y$ (hip y), $z$ (hip z),
$a$ (hip angle), $b$ (knee angle),
hip center, hip range,
knee center, knee range,
hip size, knee size,
foot center, foot range, foot size.

Existing designs and their performance scores:

\texttt{|design-start|[+001.250, +009.000, \ldots, -001.400, +000.180]|design-end|}
\texttt{|label-start|+000.703|label-end|}

\texttt{|design-start|[+002.100, +008.500, \ldots, -000.950, +000.260]|design-end|}
\texttt{|label-start|+000.729|label-end|}

Please propose one new morphology design to maximize the score.

Format: each value must be a signed float with three decimal places (e.g., \texttt{+000.123}, \texttt{-001.500}).

Response:

\texttt{|design-start|[+001.800, +009.250, \ldots, -001.200, +000.220]|design-end|}
\end{tcolorbox}

\subsection{
\textcolor{black}{Effect of High-Dimensional GP Priors}
}
\label{app: high-dim}

% \begin{table*}[t]
% \centering
% \small
% \setlength{\tabcolsep}{8pt}
% \renewcommand{\arraystretch}{1.15}

% \caption{
% \textbf{Comparison between different kernel.}
% }

% \label{tab:ab-kernel}

% \begin{tabular}{lcccc}

% \toprule
% Method & Ant Morphology & D’Kitty Morphology & TF Bind~8 & TF Bind~10 \\
% \midrule
% Grad-mean & $0.351\pm 0.002$ & $0.879\pm 0.015$ & $0.623 \pm 0.030$ & $0.394\pm 0.004$ \\
% Grad (Mean) & $0.472 \pm 0.013$  & $ \pm $ & $0.441 \pm 0.008$ & $ \pm $ \\
% \midrule 
% Grad-mean & $0.644\pm 0.039$ & $0.907\pm 0.016$ & $0.666\pm 0.011$ & $0.695\pm 0.027$ \\
% Grad (Mean) & $0.709 \pm 0.002$  & $ \pm $ & $0.843 \pm 0.082$ & $ \pm $ \\

% \midrule
% Grad-EI & $0.368\pm 0.046$ & $0.872\pm 0.004$ & $0.595 \pm 0.019$ & $0.453\pm 0.047$  \\
% Grad (EI) & $0.504 \pm 0.013$ & $ \pm $ & $0.436 \pm 0.004$  & $ \pm $ \\
% \midrule
% Grad-EI & $0.626\pm 0.002$ & $0.901\pm 0.045$ & $0.673\pm 0.012$ & $0.689\pm 0.013$ \\
% Grad (EI) & $0.655 \pm 0.002$ & $ \pm $ & $0.864 \pm 0.091$  & $ \pm $ \\

% % \midrule
% % Difference & $-0.001$ & $+0.004$ & $-0.005$  & $+0.012$ \\

% \bottomrule
% \end{tabular}

% \end{table*}

\begin{table}[H]
\centering
\small
\setlength{\tabcolsep}{6pt}
\renewcommand{\arraystretch}{1.15}

\caption{
\textcolor{black}{
\textbf{Effect of high-dimensional GP kernels on baseline performance.}
% We replace the standard RBF kernel by the high-dimensional kernel with dimensionality-scaled prior proposed by Hvarfner et al. (ICML 2024).
We compare a plain RBF GP with the high-dimensional formulation of \citet{hvarfner2024vanilla},
which incorporates dimensionality-scaled priors (DSP) for stable modeling in high-dimensional settings.
We report both the maximum and median scores, and use $\Delta$ to denote the change of performance from plain RBF to the high-dimensional configuration.
% Even with this stronger kernel, our method (DiBO) remains
% consistently superior across all tasks.
}
}

\label{tab:hd_kernel}

\begin{subtable}{\textwidth}
\setlength{\tabcolsep}{6pt}
\centering
\caption{Maximum Scores}
\begin{tabular}{llcccc}

\toprule
Method & Kernel & Ant Morphology & D'Kitty Morphology & TF Bind~$8$ & TF Bind~$10$ \\
\midrule

\multirow{3}{*}{Grad-mean}
& RBF
& $0.644\pm 0.039$ 
& $0.907\pm 0.016$ 
& $0.666\pm 0.011$ 
& $0.695\pm 0.027$ \\

& High-dim
& $0.709 \pm 0.002$  
& $0.920 \pm 0.008$ 
& $0.843 \pm 0.082$ 
& $0.736 \pm 0.016$ \\

& $\Delta$
& $+0.065 \uparrow$
& $+0.013 \uparrow$
& $+0.177 \uparrow$
& $+0.041 \uparrow$ \\

\midrule

\multirow{3}{*}{Grad-EI}
& RBF
& $0.626\pm 0.002$ 
& $0.901\pm 0.045$ 
& $0.673\pm 0.012$ 
& $0.689\pm 0.013$ \\

& High-dim
& $0.655 \pm 0.002$ 
& $0.923 \pm 0.010$ 
& $0.864 \pm 0.091$  
& $0.727 \pm 0.024$ \\

& $\Delta$
& $+0.029 \uparrow$
& $+0.022 \uparrow$
& $+0.191 \uparrow$
& $+0.038 \uparrow$ \\

\midrule

\textbf{DiBO$_{\mathrm{(ours)}}$}
& --
& $0.932 \pm 0.022$ 
& $0.912 \pm 0.017$ 
& $0.946 \pm 0.043$   
& $0.741 \pm 0.027$ \\

\bottomrule
\end{tabular}
\end{subtable}

\vspace{0.8em}

\begin{subtable}{\textwidth}
\setlength{\tabcolsep}{6pt}
\centering
\caption{Median Scores}
\begin{tabular}{llcccc}

\toprule
Method & Kernel & Ant Morphology & D'Kitty Morphology & TF Bind~$8$ & TF Bind~$10$ \\
\midrule

\multirow{3}{*}{Grad-mean}
& RBF
& $0.351\pm 0.002$ 
& $0.879\pm 0.015$ 
& $0.623 \pm 0.030$ 
& $0.394\pm 0.004$ \\

& High-dim
& $0.472 \pm 0.013$  
& $0.858 \pm 0.003$ 
& $0.441 \pm 0.008$ 
& $0.461 \pm 0.008$ \\

& $\Delta$
& $+0.121 \uparrow$
& $-0.021 \downarrow$
& $-0.182 \downarrow$
& $+0.067 \uparrow$ \\

\midrule

\multirow{3}{*}{Grad-EI}
& RBF
& $0.368\pm 0.046$ 
& $0.872\pm 0.004$ 
& $0.595 \pm 0.019$ 
& $0.453\pm 0.047$  \\

& High-dim
& $0.504 \pm 0.013$ 
& $0.849 \pm 0.008$ 
& $0.436 \pm 0.004$  
& $0.459 \pm 0.024$ \\

& $\Delta$
& $+0.136 \uparrow$
& $-0.023 \downarrow$
& $-0.159 \downarrow$
& $+0.006 \uparrow$ \\

\midrule

\textbf{DiBO$_{\mathrm{(ours)}}$}
& --
& $0.524 \pm 0.038$ 
& $0.775 \pm 0.010$ 
& $0.473 \pm 0.020$
& $0.491 \pm 0.006$ \\
        
\bottomrule
\end{tabular}
\end{subtable}

\end{table}

% \textcolor{black}{
% High-dimensional structured optimization is a recurring challenge across a wide range of domains, where the search space is often combinatorial, constrained, or governed by complex latent dependencies. 
% Recent black-box and model-based optimization methods have studied such challenges in sequence, molecule, protein, and structured design problems~\citep{chen2023bidirectional, chen2023parallel, yuan2023importance, yuan2024design, yuan2025paretoflow, yuan2026diffusion, spade}. 
% \citep{chen2023bidirectional, chen2023parallel, yuan2024design, yuan2025paretoflow, spade}. 
% Beyond classical design optimization, related high-dimensional optimization, decision-making, and representation problems also appear in large language model ~\citep{zhu2025soft, yu2025git, xiang2025fine, zhang2023towards, du2026optimizing, zhang2026preference}, 3D and multimodal understanding~\citep{nardi2017algorithmic, lascomp, li2025multi, 4dpc2hat, lee2024multimodal}, and mobile and video streaming scenarios ~\citep{liu2023sparse, benzaghta2025data, feng2020high, hu2025livevv, duan2025semconf, kang2024bridging}. 
% This broader context highlights the importance of carefully evaluating optimization methods under high-dimensional assumptions.
% }

\textcolor{black}{
In Bayesian optimization, GP surrogates provide a standard way to model unknown black-box objectives and guide acquisition-based search~\citep{garnett_bayesoptbook_2023}. 
However, in high-dimensional settings, the effectiveness of GP-based BO can depend strongly on kernel choices and hyperparameter priors, especially the lengthscale prior~\citep{binois2022survey}. 
% High-dimensional structured optimization also appears in recent BBO \citep{chen2023bidirectional, chen2023parallel, yuan2024design, yuan2025paretoflow, spade} and multi-modal understanding studies \citep{lee2024multimodal, 4dpc2hat, zhang2026preference, du2026optimizing, lascomp}.
Given that such high-dimensional challenges are increasingly prevalent in recent BBO \citep{chen2023bidirectional, chen2023parallel, yuan2024design, yuan2025paretoflow, spade} and multi-modal studies \citep{lee2024multimodal, 4dpc2hat, zhang2026preference, du2026optimizing, lascomp}, addressing this scaling issue is critical.
Fortunately, recent work shows that a simple dimension-scaled GP prior can substantially improve vanilla BO in high-dimensional search spaces~\citep{hvarfner2024vanilla}. 
Following this line of work, we implement our GP-based baselines following BoTorch~\citep{botorch} and adopt the dimension-scaled GP prior configuration of \citet{hvarfner2024vanilla} for both \textit{Grad (mean)} and \textit{Grad (EI)}.
Readers can refer to a detailed discussion \href{https://github.com/meta-pytorch/botorch/discussions/2451}{here}.
This configuration scales the GP lengthscale prior with the input dimensionality, making the surrogate better calibrated for high-dimensional design spaces. 
All other components of the baselines are kept fixed.
}

\textcolor{black}{
To justify this choice, we further compare the adopted high-dimensional GP configuration with a plain RBF GP configuration in Table~\ref{tab:hd_kernel}. 
The high-dimensional configuration consistently improves the GP-based baselines in most maximum-score comparisons and in several median-score comparisons, empirically confirming that GP prior calibration is an important implementation detail in high-dimensional design spaces. 
These results also show that our main conclusions remain robust under a stronger and more carefully calibrated GP-based baseline configuration.
}

% \caption{
% Effect of high-dimensional GP priors on GP-based baselines. 
% We compare the original RBF GP configuration with the dimension-scaled GP prior configuration of \citet{hvarfner2024vanilla} for Grad-mean and Grad-EI. 
% All other experimental settings are kept unchanged. 
% $\Delta$ denotes the change from RBF to the high-dimensional configuration. 
% Even with this stronger GP baseline, DiBO remains competitive and achieves the best or near-best performance in most cases.
% }

\subsection{Median Results}
\label{appendix: median-results}

This section reports the median ($50$-th percentile) normalized oracle scores on all Design-Bench tasks.
Consistent with the main results in Table~\ref{tab: main-table-max}, DiBO achieves state-of-the-art or competitive median performance across tasks.

\begin{table}[ht]
	\centering
	\caption{
    \textbf{Experimental results in $50$-th percentile normalized scores on four tasks for comparison.}
    Results are averaged over $8$ random seeds (mean ± std).
    Rank Mean and Rank Median denote the average and median rank across all tasks, respectively.
    The highest and second-highest scores per task are highlighted in \greentag{green} and \bluetag{blue}, respectively.
    }  
	\label{tab: continuous_median}
        \resizebox{\textwidth}{!}{%
	\begin{tabular}{ccccc|ccc}
		\toprule
		\textbf{Method} & \textbf{Ant Morphology} & \textbf{D’Kitty Morphology} & \textbf{TF Bind $8$} & \textbf{TF Bind $10$} & \textbf{Mean Score $\uparrow$} & \textbf{Rank Mean $\downarrow$} & \textbf{Rank Median $\downarrow$}\\
		\midrule
		$\mathcal{D}$(\textbf{best}) & $0.565$ & $0.884$ & $0.439$& $0.511$ & $-$ & $-$ & $-$ \\
		\midrule
        % \multicolumn{7}{l}{\textit{Forward Methods}} \\
        % \cmidrule{1-7}
        % Grad-mean & $0.351\pm 0.002$ & $0.879\pm 0.015$ & \best{0.623 \pm 0.030} & $0.394\pm 0.004$ & $0.562 \pm 0.013$ & $6.5$ & $6.5$ \\
        Grad-mean & $0.472 \pm 0.013$  & $0.858 \pm 0.003$ & \second{0.441 $\pm$ 0.008} & $0.461 \pm 0.008$ & $0.558 \pm 0.008$ & \best{3.5} & \second{3.5} \\
        % Grad-EI & $0.368\pm 0.046$ & $0.872\pm 0.004$ & $0.595 \pm 0.019$ & $0.453\pm 0.047$ & \second{0.572 $\pm$ 0.029} & $6.0$ & $5.0$ \\
        Grad-EI & \second{0.504 $\pm$ 0.013} & $0.849 \pm 0.008$ & $0.436 \pm 0.004$  & $0.459 \pm 0.024$ & \second{0.562 $\pm$ 0.012} & \second{4.0} & $4.0$ \\
        COMs & $0.375\pm 0.046$ & \best{0.887\pm 0.007} & $0.399\pm 0.048$ & \second{0.487 $\pm$ 0.039} & $0.537 \pm 0.035$ & $5.25$ & $5.0$ \\
        ICT & $0.407\pm 0.022$ & $0.863\pm 0.049$ & $0.366\pm 0.010$ & $0.365\pm 0.017$ & $0.500 \pm 0.025$ & $9.25$ & $9.0$ \\
        MATCH-OPT & $0.435\pm 0.049$ & $0.875\pm 0.022$ & $0.431\pm 0.034$ & $0.373\pm 0.007$ & $0.529 \pm 0.028$ & $6.25$ & $4.5$ \\
        UniSO-T & $0.405\pm 0.026$ & $0.842\pm 0.043$ & $0.418\pm 0.013$ & $0.431\pm 0.015$ & $0.524 \pm 0.024$ & $7.25$ & $7.0$ \\
		\midrule
        % \multicolumn{7}{l}{\textit{Inverse Methods}} \\
        % \cmidrule{1-7}
        CbAS & $0.302\pm 0.033$ & $0.773\pm 0.047$ & $0.394\pm 0.009$ & $0.373\pm 0.009$ & $0.461 \pm 0.025$ & $13.0$ & $13.5$ \\
        ExPT & $0.465 \pm 0.039$ & $0.792\pm 0.016$ & $0.355\pm 0.011$ & $0.356\pm 0.027$ & $0.492 \pm 0.023$ & $11.5$ & $13.0$ \\
        MIN & $0.370\pm 0.002$ & $0.841\pm 0.045$ & $0.282\pm 0.012$ & $0.386\pm 0.013$ & $0.470 \pm 0.018$ & $11.75$ & $11.5$ \\
        BONET & $0.391\pm 0.020$ & \second{0.885 $\pm$ 0.008} & $0.408\pm 0.046$ & $0.474\pm 0.025$ & $0.540 \pm 0.025$ & $4.75$ & $4.5$ \\
        OPRO & $0.380\pm 0.045$ & $0.829\pm 0.037$ & $0.407\pm 0.050$ & $0.432\pm 0.014$ & $0.512 \pm 0.037$ & $9.0$ & $9.0$ \\
        GTG & $0.372\pm 0.024$ & $0.765\pm 0.025$ & $0.360\pm 0.008$ & $0.382\pm 0.034$ & $0.470 \pm 0.023$ & $12.75$ & $12.5$ \\
        DDOM & $0.323\pm 0.045$ & $0.835\pm 0.007$ & $0.373\pm 0.027$ & $0.454\pm 0.017$ & $0.496 \pm 0.024$ & $10.0$ & $10.0$ \\
		\midrule
        CMA-ES & $0.335\pm 0.019$ & $0.686\pm 0.035$ & $0.388\pm 0.028$ & $0.436\pm 0.005$ & $0.461 \pm 0.022$ & $11.75$ & $11.5$ \\
        MCTS-transfer & $0.323\pm 0.049$ & $0.832\pm 0.041$ & $0.310\pm 0.044$ & $0.453 \pm 0.022$ & $0.480 \pm 0.039$ & $11.5$ & $12.0$ \\
		\midrule
		% \textbf{DiBO}$_{\mathrm{(ours)}}$ & \best{0.524 \pm 0.015} & $0.837 \pm 0.003$ & \best{0.473 \pm 0.061} & \best{0.504 \pm 0.012} & \best{0.589 \pm 0.026} & \best{3.0} & \best{1.0} \\
        \textbf{DiBO}$_{\mathrm{(ours)}}$ & \best{0.524 \pm 0.038} & $0.775 \pm 0.010$ & \best{0.473 \pm 0.020} & \best{0.491 \pm 0.006} & \best{0.566 \pm 0.019} & \second{4.0} & \best{1.0} \\
		\bottomrule
	\end{tabular}
	}
\end{table}

% \subsection{
% \textcolor{red}{Comparison with Inference-Only dLLM + MCTS}
% }
% \label{appendix:dllm-mcts}

% \ZP{This section has been moved to main text.}

% We additionally consider the inference-only diffusion LLM approach combined with Monte Carlo Tree Search (MCTS)~\citep{yuan2026diffusion}.
% This method performs optimization by querying a frozen diffusion language model and does not involve any form of domain adaptation or post-training.
% Note that this approach is designed to operate in an extremely few-shot optimization regime (e.g. $n_{pool} = 10$) in stead of ours ($n_{pool} = 500$).
% Due to this difference in problem setting and optimization regime, we do not include this method in the main quantitative comparisons.
% We refer interested readers to~\citet{yuan2026diffusion} for a detailed evaluation of this approach
% under its intended few-shot setting.

\subsection{Additional Evaluation on Top-$K$ Metrics and RNA Design Tasks}
\label{app:rna_topk}
\textcolor{black}{
We provide additional evaluation along two dimensions.
First, beyond the $100$-th percentile score used in the main results, we report Top-$K$ average scores, where Top-$K$ denotes the average oracle score among the top $K$ generated candidates out of $128$ candidates.
We report Top-$5$, Top-$10$, Top-$20$, and the $50$-th percentile score to characterize not only the best generated candidate but also the quality of the generated candidate set. 
}

\textcolor{black}{
As shown in Table~\ref{tab:ab-topk}, DiBO remains strong on the original Design-Bench tasks under Top-$K$ and median metrics.
In particular, it achieves the best scores across all reported metrics on Ant Morphology, TF Bind $8$, and TF Bind $10$, while remaining competitive on D'Kitty Morphology as discussed.
}

\textcolor{black}{
Second, we extend the evaluation to three RNA optimization tasks, RNA-A/B/C, adapted from BootGen~\citep{kim2023bootstrapped}.
These tasks require designing length-$14$ RNA sequences to maximize transcription factor binding activity, providing an additional biological sequence domain beyond the Design-Bench tasks.
We compare DiBO with COMs and ExPT, the two representative baselines from our main evaluation.
}
\textcolor{black}{
The results are comparable but more mixed on the RNA tasks, as shown in Table~\ref{tab:ab-rna}.
DiBO achieves the best $100$-th percentile score on RNA-A and RNA-C, indicating that it can still discover high-scoring candidates in these additional biological sequence tasks.
%However, its Top-$K$ and 50-th percentile scores are less competitive, especially on RNA-B, suggesting that the generated candidate distribution is less consistently concentrated in high-score regions.
}
% One possible reason is that the $\mathcal{D}$(\textbf{best}) of these RNA tasks are extremely low (around 0.12 compared to the max possible scored set as 1.0), means the best point we have is bnot that informative. This can make learning stable local improvements from prompt-response pairs more difficult.
% By contrast, COMs and ExPT directly optimize a conservative surrogate or a learned generative latent model, which may yield more concentrated candidate sets on these RNA tasks.

\begin{table}[H]
\centering
\small
\setlength{\tabcolsep}{8pt}
\renewcommand{\arraystretch}{1.15}

\caption{
\textbf{Additional evaluation on Top-$K$ metrics on Design-Bench tasks.}
% \textcolor{black}{
We report the $100$-th percentile, Top-$5$/$10$/$20$ average, and $50$-th percentile normalized oracle scores over $128$ generated candidates.
The highest and second-highest scores per task are highlighted in \greentag{green} and \bluetag{blue}, respectively.}
% }
\label{tab:ab-topk}

\begin{tabular}{ll|cc|c}
\toprule
Task & Metric & COMs & ExPT & \textbf{DiBO$_{\mathrm{(ours)}}$} \\
\midrule

\multirow{5}{*}{Ant Morphology}
& $100$-th  & $0.647 \pm 0.020$ & \second{0.929 $\pm$ 0.049} & \best{0.932 \pm 0.022} \\
& Top-$5$   & $0.628 \pm 0.014$ & \second{0.904 $\pm$ 0.028} & \best{0.906 \pm 0.012} \\
& Top-$10$  & $0.621 \pm 0.020$ & \second{0.863 $\pm$ 0.041} & \best{0.880 \pm 0.010} \\
& Top-$20$  & $0.592 \pm 0.012$ & \second{0.828 $\pm$ 0.052} & \best{0.847 \pm 0.011} \\
& $50$-th   & $0.375 \pm 0.046$ & \second{0.465 $\pm$ 0.039} & \best{0.524 \pm 0.038} \\

\midrule

\multirow{5}{*}{D’Kitty Morphology}
& $100$-th  & \second{0.934 $\pm$ 0.008} & \best{0.950 \pm 0.041} & $0.912 \pm 0.017$ \\
& Top-$5$   & \second{0.912 $\pm$ 0.005} & \best{0.935 \pm 0.020} & $0.900 \pm 0.006$ \\
& Top-$10$  & \second{0.908 $\pm$ 0.005} & \best{0.918 \pm 0.009} & $0.894 \pm 0.006$ \\
& Top-$20$  & \best{0.901 \pm 0.010} & \second{0.892 $\pm$ 0.028} & $0.884 \pm 0.005$ \\
& $50$-th   & \best{0.887 \pm 0.007} & \second{0.792 $\pm$ 0.016} & $0.775 \pm 0.010$ \\

\midrule

\multirow{5}{*}{TF Bind~$8$}
& $100$-th  & \second{0.843 $\pm$ 0.046} & $0.810 \pm 0.044$ & \best{0.946 \pm 0.043} \\
& Top-$5$   & \second{0.824 $\pm$ 0.053} & $0.790 \pm 0.021$ & \best{0.879 \pm 0.011} \\
& Top-$10$  & \second{0.742 $\pm$ 0.042} & $0.733 \pm 0.058$ & \best{0.832 \pm 0.018} \\
& Top-$20$  & \second{0.701 $\pm$ 0.022} & $0.696 \pm 0.062$ & \best{0.766 \pm 0.016} \\
& $50$-th   & \second{0.399 $\pm$ 0.048} & $0.355 \pm 0.011$ & \best{0.473 \pm 0.020} \\

\midrule

\multirow{5}{*}{TF Bind~$10$}
& $100$-th  & \second{0.709 $\pm$ 0.025} & $0.703 \pm 0.022$ & \best{0.741 \pm 0.027} \\
& Top-$5$   & $0.626 \pm 0.040$ & \second{0.652 $\pm$ 0.012} & \best{0.681 \pm 0.021} \\
& Top-$10$  & $0.608 \pm 0.016$ & \second{0.613 $\pm$ 0.011} & \best{0.652 \pm 0.017} \\
& Top-$20$  & $0.588 \pm 0.013$ & \second{0.596 $\pm$ 0.008} & \best{0.621 \pm 0.009} \\
& $50$-th   & \second{0.487 $\pm$ 0.039} & $0.356 \pm 0.027$ & \best{0.491 \pm 0.006} \\

\bottomrule
\end{tabular}

\end{table}

\begin{table}[H]
\centering
\small
\setlength{\tabcolsep}{8pt}
\renewcommand{\arraystretch}{1.15}

\caption{
\textbf{Additional evaluation on RNA design tasks.}
\textcolor{black}{
We extend our evaluation to three RNA optimization tasks (RNA-A/B/C) following the BootGen \citep{kim2023bootstrapped}. 
These tasks involve designing length-$14$ RNA sequences to maximize transcription factor binding activity. 
Results report the normalized oracle score under different Top-$K$ metrics, where Top-$K$ denotes the average score of the top K generated candidates. 
% We compare against two strong baselines COMs and ExPT in our original Design-Bench evaluation.
We compare against two representative strong baselines, COMs and ExPT, from our original Design-Bench evaluation.
The highest and second-highest scores per task are highlighted in \greentag{green} and \bluetag{blue}, respectively.
}
}
\label{tab:ab-rna}

\begin{tabular}{ll|cc|c}
\toprule
Task & Metric & COMs & ExPT & \textbf{DiBO$_{\mathrm{(ours)}}$} \\
\midrule

\multirow{5}{*}{RNA-A}
& $100$-th  & $0.362 \pm 0.011$ & \second{0.365 $\pm$ 0.031} & \best{0.395 \pm 0.015} \\
& Top-$5$ & $0.322 \pm 0.024$ & \second{0.331 $\pm$ 0.014} & \best{0.335 \pm 0.014} \\
& Top-$10$  & \second{0.305 $\pm$ 0.020} & \best{0.308 \pm 0.016} & $0.258 \pm 0.012$ \\
& Top-$20$  & \second{0.264 $\pm$ 0.017} & \best{0.294 \pm 0.025} & $0.224 \pm 0.007$ \\
& $50$-th  & \best{0.155 \pm 0.023} & $0.133 \pm 0.019$ & \second{0.141 $\pm$ 0.004} \\

\midrule

\multirow{5}{*}{RNA-B}
& $100$-th  & \best{0.350 \pm 0.063} & \second{0.348 $\pm$ 0.011} & $0.299 \pm 0.017$ \\
& Top-$5$   & \best{0.332 \pm 0.023} & \second{0.313 $\pm$ 0.007} & $0.229 \pm 0.009$ \\
& Top-$10$  & \best{0.299 \pm 0.030} & \second{0.292 $\pm$ 0.006} & $0.203 \pm 0.009$ \\
& Top-$20$  & \best{0.290 \pm 0.014} & \second{0.275 $\pm$ 0.003} & $0.177 \pm 0.059$ \\
& $50$-th   & \best{0.162 \pm 0.028} & \second{0.138 $\pm$ 0.009} & $0.087 \pm 0.004$ \\

\midrule

\multirow{5}{*}{RNA-C}
& $100$-th  & $0.445 \pm 0.018$ & \second{0.502 $\pm$ 0.016} & \best{0.528 \pm 0.037} \\
& Top-$5$   & $0.403 \pm 0.009$ & \best{0.487 \pm 0.013} & \second{0.413 $\pm$ 0.015} \\
& Top-$10$  & \second{0.392 $\pm$ 0.031} & \best{0.430 \pm 0.004} & $0.322 \pm 0.013$ \\
& Top-$20$  & \second{0.359 $\pm$ 0.006} & \best{0.415 \pm 0.007} & $0.280 \pm 0.009$ \\
& $50$-th   & \second{0.205 $\pm$ 0.044} & \best{0.247 \pm 0.023} & $0.162 \pm 0.007$ \\

\bottomrule
\end{tabular}

\end{table}

% \textcolor{red}{
% One possible reason is that the offline pools for these RNA tasks provide weak high-score anchors: the best observed offline scores, denoted as $\mathcal{D}_{\mathrm{best}}$, are only around $0.12$ under normalization, far from the maximum possible score of $1.0$.
% In this setting, prompt-response pairs mainly capture relative improvements within a low-score region, which may provide limited signal for learning stable local improvements toward truly high-scoring RNA sequences.
% Through context resampling and generative diversity, DiBO can still occasionally identify strong candidates, leading to competitive or best 100-th percentile scores.
% Nevertheless, shifting the entire candidate distribution so that many generated samples are consistently high-scoring is more challenging.
% }

% \textcolor{red}{
% By contrast, COMs and ExPT use optimization mechanisms that may better concentrate candidate sets under weak offline anchors.
% COMs directly optimizes a conservative surrogate objective designed to reduce overestimation on out-of-distribution inputs.
% ExPT uses synthetic pretraining to learn a few-shot experimental-design prior over candidate optimization.
% These mechanisms may yield more stable Top-$K$ and median performance on the RNA tasks.
% Overall, the RNA results suggest that DiBO is effective at best-candidate discovery in this additional biological domain, but improving distribution-level concentration under weak offline supervision remains an important direction.
% }

\subsection{
\textcolor{black}{
Pretraining-Overlap Audit via Ranking Probing
}
}
\label{app:pretrain_audit}

\textcolor{black}{
Since DiBO is initialized from a pretrained diffusion LLM, a natural concern is whether the pretrained checkpoint already contains task-specific design-label knowledge from public benchmark assets.
Because the pretraining corpus of \textsc{LLaDA-8B-Instruct} is \href{https://github.com/ML-GSAI/LLaDA\#pre-training-and-supervised-fine-tuning}{not publicly available}, an exact dataset-level overlap audit is not possible.
We therefore perform a behavioral audit that probes whether the pretrained model can rank candidate designs according to their ground-truth labels before any domain adaptation or post-training.
}

\textcolor{black}{
For each task and each ranking size $K \in \{2,3,4,5\}$, we sample $500$ groups of $K$ candidate designs from the offline pool.
The pretrained model is given only the rendered designs, without their labels, and is prompted to rank them from highest score to lowest score.
The following probing prompt shows an example when $K=3$, where each \texttt{<design>} is rendered in the same format as the corresponding benchmark task.
\begin{center}
\vspace{-0.4em}
\setlength{\fboxsep}{4pt}
\fbox{
\begin{minipage}{0.5\linewidth}
\small
You are given $3$ candidate designs.\\
Rank them from highest score to lowest score.\\
Return only a ranking over the option letters and no other text.\\[0.4em]
A: \texttt{<design A>}\\
B: \texttt{<design B>}\\
C: \texttt{<design C>}\\[0.4em]
Answer:
\end{minipage}
}
\vspace{-0.4em}
\end{center}
}

% For each group, we enumerate all $K!$ possible answer completions, such as \texttt{A > B > C}, \texttt{A > C > B}, and \texttt{B > A > C} when $K=3$.
% Each completion is scored by pseudo log-likelihood under the pretrained model.
% Specifically, for a completion $a$, we concatenate it with the probing prompt, mask one answer token at a time, and sum the conditional log-probabilities of the original answer tokens:
% \[
% \mathrm{PLL}(a)
% =
% \sum_{j}
% \log p_{\theta}(a_j \mid \mathrm{prompt}, a_{\setminus j}) .
% \]
% We then sort all completions by this score and record the rank of the ground-truth ordering determined by the oracle labels.
% If the pretrained model already encodes benchmark-specific design-label knowledge, the correct ordering should be ranked close to first; under random ranking, its expected rank is $(K!+1)/2$.

\textcolor{black}{
For each group, we enumerate all $K!$ possible answer rankings, such as \texttt{A > B > C}, \texttt{A > C > B}, and \texttt{B > A > C} when $K=3$.
We then compute the log probability assigned by the pretrained model to each possible ranking and sort all rankings by their model scores.
Finally, we record the rank of the ground-truth ordering determined by the oracle labels.
If the pretrained model already encodes benchmark-specific design-label knowledge, the correct ordering should be ranked close to $1$; under random ranking, its expected rank is $(K!+1)/2$ (e.g., $3.5$ when $K=3$).
}

\begin{table*}[ht]
\centering
\small
\setlength{\tabcolsep}{8pt}
\renewcommand{\arraystretch}{1.15}

% \caption{
% \textcolor{red}{
% \textbf{Pretraining-overlap audit via ranking probing.} 
% We test whether the pretrained LLaDA encodes task-specific design–label ranking knowledge before any domain adaptation or post-training. 
% For each task and each ranking size $K \in \{2,3,4,5\}$, we sample 500 groups of $K$ designs from the offline pool. 
% For each group, the ground-truth ranking is determined by their labels, and all possible permutations are enumerated and scored by the pretrained model. 
% We report the average rank assigned to the ground-truth ordering. 
% The “Random Baseline” corresponds to the expected rank under uniform random ranking.
% }
% }
\caption{\textbf{Pretraining-overlap audit via ranking probing.}
\textcolor{black}{
We test whether the pretrained \textsc{LLaDA-8B-Instruct} checkpoint encodes task-specific design-label ranking knowledge before any domain adaptation or post-training.
For each task and each $K \in \{2,3,4,5\}$, we sample $500$ groups of $K$ designs, enumerate all $K!$ possible rankings, and score each ranking by the pretrained model.
We report the average rank assigned to the ground-truth ordering, where lower is better.
The random baseline is $(K!+1)/2$.
}
}
\label{tab:ab-rank}

\begin{tabular}{lcccc}

\toprule
Task & $K=2$ & $K=3$ & $K=4$ & $K=5$ \\
\midrule
Random Guess & $1.50$ & $3.50$ & $12.5$ & $60.5$ \\
\midrule

Ant Morphology      & $1.55$ & $3.39$ & $11.77$ & $64.40$ \\
D’Kitty Morphology  & $1.55$ & $3.30$ & $11.84$ & $60.90$ \\
TF Bind~8           & $1.56$ & $3.43$ & $11.62$ & $61.96$ \\
TF Bind~10          & $1.58$ & $3.29$ & $11.62$ & $63.00$ \\
RNA-A               & $1.59$ & $3.49$ & $11.85$ & $62.94$ \\
RNA-B               & $1.56$ & $3.50$ & $11.51$ & $63.54$ \\
RNA-C               & $1.54$ & $3.52$ & $11.58$ & $64.44$ \\

\bottomrule

\end{tabular}

% \begin{tabular}{lcccc}

% \toprule
% Task & K=2 & K=3 & K=4 & K=5 \\
% \midrule
% Random Guess & 1.50 & 3.50 & 12.5 & 60.5 \\
% \midrule

% Ant Morphology      & 1.55 & 3.39 & 11.77 & 64.40 \\
% D’Kitty Morphology  & 1.55 & 3.30 & 11.84 & 60.90 \\
% TF Bind~8           & 1.56 & 3.43 & 11.62 & 61.96 \\
% TF Bind~10          & 1.58 & 3.29 & 11.62 & 63.00 \\
% RNA-A               & 1.59 & 3.49 & 11.85 & 62.94 \\
% RNA-B               & 1.56 & 3.50 & 11.51 & 63.54 \\
% RNA-C               & 1.54 & 3.52 & 11.58 & 64.44 \\

% \bottomrule

% \end{tabular}

\end{table*}

\textcolor{black}{
As shown in Table~\ref{tab:ab-rank}, the average rank of the correct ordering is close to the random baseline across tasks and values of $K$.
This suggests that the pretrained checkpoint does not exhibit meaningful prior knowledge of the design-label relationship.
}

\end{document}